\documentclass[reprint,
groupedaddress,
showpacs,
nofootinbib,
prd,
aps,
floatfix,
]{revtex4-2}

\usepackage{amsmath,amssymb,amsfonts,amsthm}
\usepackage[english]{babel}
\usepackage{url}
\usepackage{bm}
\usepackage[latin1]{inputenc}
\usepackage{graphicx,rotating}
\usepackage[colorlinks=true,linkcolor=black, citecolor=blue]{hyperref}
\usepackage{csvsimple}
\usepackage{booktabs}
\usepackage{multirow, array}
\usepackage{slashed}
\usepackage{xcolor}
\usepackage[export]{adjustbox}
\usepackage{dcolumn}

\usepackage{natbib}

\usepackage[normalem]{ulem}

\newcommand{\dd}{\mathrm{d}}

\newcommand{\ee}{E_e}
\newcommand{\enu}{E_\nu}
\newcommand{\pe}{p_e}

\begin{document}
\title{$\beta$ Recoil spectroscopy as a Beyond Standard Model laboratory}
\author{Leendert Hayen}
\email[Corresponding author: ]{hayen@lpccaen.in2p3.fr}
\affiliation{LPC Caen, ENSICAEN, Universit\'e de Caen, CNRS/IN2P3, Caen, France}

\date{\today}
\begin{abstract}
We consider the nuclear recoil spectrum following allowed $\beta$ decay as a laboratory for Beyond Standard Model searches, and perform an exhaustive treatment of the Standard Model prediction. We consider final state interactions, nuclear structure corrections, kinematic corrections and radiative corrections, derive new results and show that the nuclear recoil spectrum is theoretically exceptionally clean. We construct a clean statistical framework and analyze the Fisher information for extracting exotic current constraints and testing Cabibbo-Kobayashi-Maskawa top-row unitarity, and consider a number of sterile neutrino scenarios. In addition, we find that for select isotopes separate measurement of the $\beta$-asymmetry and recoil spectrum can achieve an increase in the statistical power of up to a factor 40.
\end{abstract}

\maketitle


\section{Introduction}

The study of nuclear recoils following $\beta$ decay is almost as old as $\beta$ decay itself, and tracks the major open questions in the field and technological advances over the decades. Starting in 1936, Leipunski \cite{Leipunski1936} reports on the detection of $^{11}$C recoils in search of missing energy deemed to originate from the neutrino. Emerging either as charged recoiling particles from thin depositions \cite{Kofoed-Hansen1954} or localized droplet formation in specialised cloud chambers \cite{Crane1938, Crane1939}, one may already recognize the first signs of a Dalitz distribution. Theoretically, the seminal paper by Kofoed-Hansen \cite{Kofoed-Hansen1948} and works by Tolhoek and de Groot \cite{deGroot1950, Tolhoek1951b, Tolhoek1951c, Tolhoek1951d} set the stage for the following decade and the role of nuclear recoil spectroscopy in it. As a result, following strong signs for existence of the neutrino \cite{Allen1948, Allen1948} before its direct detection by Reines and Cowan \cite{Reines1953, Reines1960}, the search turns towards the nature of the weak interaction. Before the discovery of parity-violation \cite{Wu1957, Wu1959}, the recoiling energy spectrum is recognised as a sensitive probe of the Lorentz structure of the $\beta$-decay interaction due to its large change in the predicted $\beta$-$\nu$ angular correlation. With the advent of primitive nuclear reactor piles (allowing for large neutron irradiation of, e.g., Beryllium powders), the $\beta^-$ decay of $^6$He becomes a quick favourite with its large recoil endpoint energy of $\sim 1.4$keV. Detailed study eventually helps settle the question of the $V$-$A$ interaction \cite{Allen1959} - despite a number of wrong turns \cite{Rustad1953, Rustad1955}. Closing out the initial period, a landmark result is published with a significant jump in precision by Johnson \textit{et al.} \cite{Johnson1963}, and a more detailed theoretical analysis by Kleppinger et al. \cite{Kleppinger1977}.

In parallel, the development of neutron beams throughout the 1950's inspired a decades-long program \cite{Robson1955, Robson1958} through the first determinations of the axial vector coupling constant, $g_A$, \cite{Dobrozemsky1957, Stratowa1978, Byrne2002, Byrne2000} up to the current generation of experiments \cite{Beck2023, Hassan2021, Gonzalez2026}. Only recently, the first Dalitz distribution of neutron $\beta$ decay was mapped in the Nab experiment \cite{Gonzalez2026} as a milestone on the road of extracting the beta-neutrino angular correlation from the proton recoil spectrum. 

With the advent of ion and atom traps one allowed for fine control of the initial state with large densities and low cloud temperatures. Several exciting results have emerged from these efforts \cite{Gorelov2000, Gorelov2005, Couratin2012, Muller2022}, but new systematic uncertainties have cropped up. In particular, charge cloud dynamics \cite{Beck2011, Porobic2015}, molecular effects \cite{Vetter2008} and scattering off trap materials and detectors \cite{Li2013} have limited the impact on Beyond Standard Model searches. 

Recently, several new methods have been demonstrated to show eV-scale sensitivity in a calorimetric or momentum-resolving fashion. Quantum sensors such as Superconducting Tunnel Junctions have demonstrated a direct energy measurement of the $^7$Li recoil following $^7$Be electron capture \cite{Fretwell2020, Friedrich2021, Smolsky2024}, while optically levitated nanospheres have demonstrated first momentum measurements of recoiling nuclei following $\alpha$ decay \cite{Carney2023}. All the while detector technologies such as Transition Edge Sensors \cite{DeLucia2024}, Metallic Magnetic Calorimeters \cite{Kempf2018} and Kinetic inductance detectors \cite{Baselmans2012, Mazin2004} have shown a stable evolution towards lower thresholds (now often tens of eV) and higher resolution \cite{Bass2024}. These are primarily deployed in dark matter experiments, but are steadily finding their way into $\beta$ decay searches.

A careful consideration of the potential and known corrections to the recoil spectrum is therefore timely. While individual efforts have focused on specific corrections, results have often been incompatible and no exhaustive treatment currently exists. In this work we tackle several problems at once: we discuss in detail a wide variety of corrections, some of which are specific to the recoil spectrum. We derive new results for several classes of corrections, and explicitly calculate the introduced bias in typical parameter extraction. We discuss several cases of Beyond Standard Model tests using only the recoil spectrum, demonstrate their competitiveness for a variety of different isotopes. Finally, we perform an extensive statistical analysis of the sensitivity to exotic tensor currents and CKM unitarity, considering both idealized and realistic scenarios with typical detector effects. We derive several useful relationships to aid in experimental design, and explicitly demonstrate the significant increase in constraining power through $\beta$-asymmetry measurements. 

\section{Tree-level expressions}

In order to more easily make the connection to the traditional literature, we will start from the tree-level expression as it is usually written down is lepton coordinates and then transform. As such, the differential decay rate is written in the usual language by Jackson, Treiman, and Wyld \cite{Jackson1957, Jackson1957a}
\begin{align}
    \frac{d\Gamma}{dE_ed\Omega_ed\Omega_\nu} &= \frac{G_F^2V_{ud}^2}{(2\pi)^5}
    \left\{1+b\frac{m_e}{E_e}+a_{\beta\nu}\frac{\bm{p}_e\cdot\bm{p}_\nu}{E_eE_\nu} \right\} \nonumber \\
    &\times F(Z, E_e)K(Z, E_e, E_0) \nonumber \\
    &\times p_eE_ep_\nu E_\nu 
    \label{eq:JTW_angular}
\end{align}
assuming no polarization of the initial state and no final state spins are observed. Here, $F(Z,E_e)$ is the traditional Fermi function, $K(Z, E_e, E_0)$ are higher-order spectral corrections discussed below \cite{Hayen2018, Hayen2019a} and $E_0$
is the $\beta$ spectrum endpoint. 
At tree level the decay is purely three-body and we have the following relationships
\begin{equation}
    \cos \theta_{\beta\nu} = \frac{\bm{p}_f^2-\bm{p}_e^2-\bm{p}_\nu^2}{2p_ep_\nu}.
\end{equation}
Allowed values for kinematics are such that $|\cos \theta_{\beta\nu}| \leq 1$, and it traces out an allowed region in the Dalitz distribution. For a finite neutrino mass, there are four extremal points summarized in Table \ref{tab:dalitz_extremal_points}. For a massless neutrino, the maximal recoil and $\beta$ particle energy points coincide.

\begin{ruledtabular}
\begin{table*}[ht]
    \centering
    \caption{Characteristic extremal points of the beta-decay Dalitz region.
    Here $\Delta=M_i-M_f$ and $Q=M_i-M_f-m_e$.}
    \label{tab:dalitz_extremal_points}
    \begin{tabular}{l|cc}
        Point
        & Electron energy $E_e$
        & Recoil kinetic energy $T_f$
        \\
        \midrule

        A: Zero recoil
        &
        $\displaystyle
        \frac{\Delta^2+m_e^2-m_\nu^2}{2\Delta}$
        &
        $\displaystyle 0$
        \\

        B: Electron at rest
        &
        $\displaystyle m_e$
        &
        $\displaystyle
        \frac{Q^2-m_\nu^2}{2(M_f+Q)}$
        \\

        C: Maximum electron energy
        &
        $\displaystyle
        \frac{
            M_i^2+m_e^2-(M_f+m_\nu)^2
        }{2M_i}$
        &
        $\displaystyle
        \frac{
            M_f\!\left(
                [M_i-(M_f+m_\nu)]^2-m_e^2
            \right)
        }{
            2M_i(M_f+m_\nu)
        }$
        \\

        D: Maximum recoil energy
        &
        $\displaystyle
        \frac{m_e}{m_e+m_\nu}
        \frac{
            M_i^2-M_f^2+(m_e+m_\nu)^2
        }{2M_i}$
        &
        $\displaystyle
        \frac{
            \Delta^2-(m_e+m_\nu)^2
        }{2M_i}$
        \end{tabular}
\end{table*}
\end{ruledtabular}

\begin{figure}
    \centering
    \includegraphics[width=\linewidth]{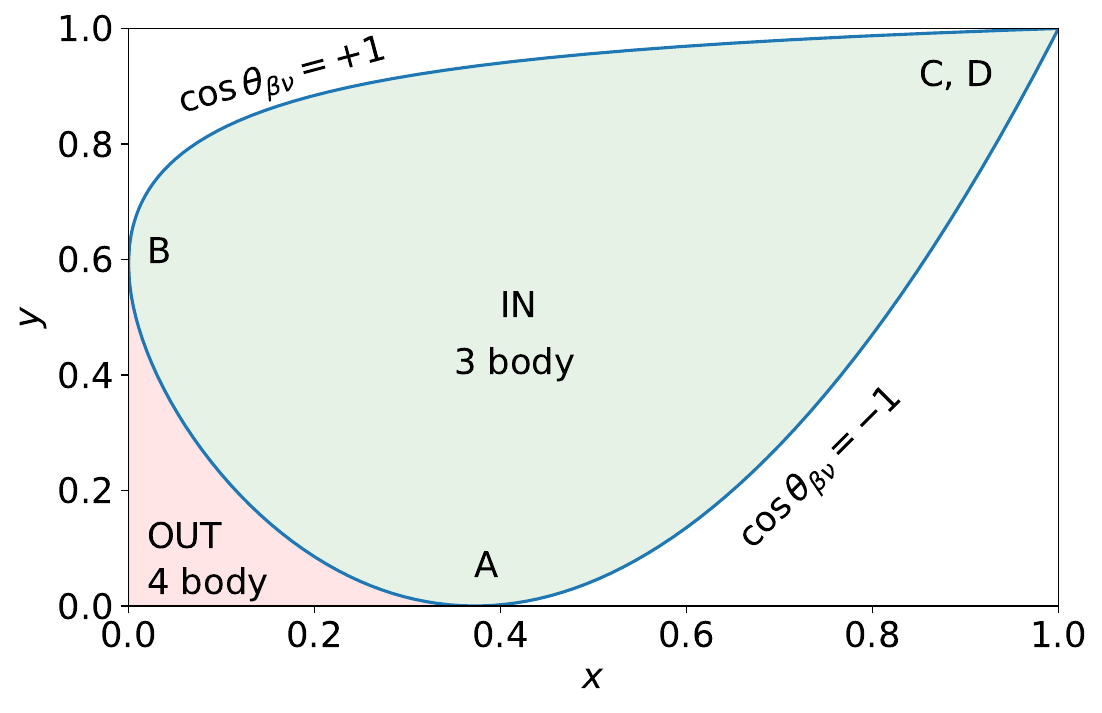}
    \caption{Typical Dalitz distribution for $\beta$ decay in terms of kinetic energies $T_e = xT_e^{max}$, $T_f = yT_f^{max}$. All three body decays and radiative decays for a vanishing proton energy are contained in the region denoted 'IN'. The 'OUT' region can only be populated by four body decays where the additional particles takes away a non-negligible amount of energy. Extremal points $A,B,C$ correspond to zero kinetic energy of the $\beta$ particle, recoil, and neutrino, respectively.}
    \label{fig:Dalitz_pedagogical}
\end{figure}

Figure \ref{fig:Dalitz_pedagogical} shows the Dalitz distribution for a massless neutrino with the three extremal points. 

In order to turn Eq. (\ref{eq:JTW_angular}) into a differential recoil spectrum use conservation of momentum and find
\begin{equation}
    d(\cos \theta_{\beta\nu}) = \frac{M}{p_ep_\nu}dE_f
    \label{eq:dcostheta}
\end{equation}
so that
\begin{align}
    \frac{d\Gamma}{dE_f} &= \frac{G_F^2V_{ud}^2}{4\pi^3} \int_{E_{e}^-}^{E_{e}^+}dE_eF(Z, E_e)K(Z, E_e, E_0)E_eE_\nu\nonumber \\
    &\times\left(1+b\frac{m_e}{E_e}+\frac{a}{2}\frac{\bm{p}_f^2-\bm{p}_e^2-\bm{p}_\nu^2}{E_eE_\nu}\right)
    \label{eq:recoil_integral}
\end{align}
where $E_{e}^{\pm}$ corresponds to $\cos\theta_{\beta\nu}(E_f, E_e) = \pm1$.  This is also often expressed as
\begin{align}
    W_0 &= \frac{G_F^2V_{ud}^2}{4\pi^3}(1+\rho^2)F(Z, E_e) \nonumber \\
    &\times \left[E_eE_\nu(1+a)+a(T_f-T_f^{max})\right]
    \label{eq:W0}
\end{align}
Throughout the rest of the manuscript we will add additional terms fo this differential amplitude, the effects of which will be discussed separately. For many of these we may use the machinery of the $\beta$ spectrum shape developed in Ref. \cite{Hayen2018, Hayen2019a}. There, many of the additional effects are defined as multiplicative factors to the zeroth-order amplitude with powers of $E_e$ and $E_0$. We may then define the following for later use
\begin{equation}
    I_{mnl}^C = \int_{E_{e,min}}^{E_{e,max}}dE_eF(Z, E_e) E_e^m(E_0-E_e)^np_e^l.
\end{equation}
where the superscript denotes the inclusion of the Coulomb interaction through the Fermi function.

In this notation, Eq. (\ref{eq:recoil_integral}) may be expressed as
\begin{align}
    \frac{d\Gamma}{dE_f} &= I_{110}^C+bm_eI^C_{010}+\frac{a}{2}(p_f^2I^C_{000}-I^C_{002}-I^C_{020}) \nonumber \\
    &\equiv I_s + bI_b + aI_a
    \label{eq:recoil_spectrum_I}
\end{align}
We may similarly define $I_{mnl} \equiv I_{mnl}^C(Z=0)$, so that
\begin{equation}
    I_{mnl}^C = I_{mnl} \pm \alpha Z \pi I_{m+1~n~l-1} + \mathcal{O}(\{\alpha Z\}^2)
    \label{eq:I_Coul_expansion}
\end{equation}
using the simplified Fermi function $F\approx 1 \pm \alpha Z \pi / \beta $ with $\beta = p_e/E_e$. Appendix \ref{app:integrals} contains a summary of closed form results for the most common integrals. 

\begin{figure}
    \centering
    \includegraphics[width=\linewidth]{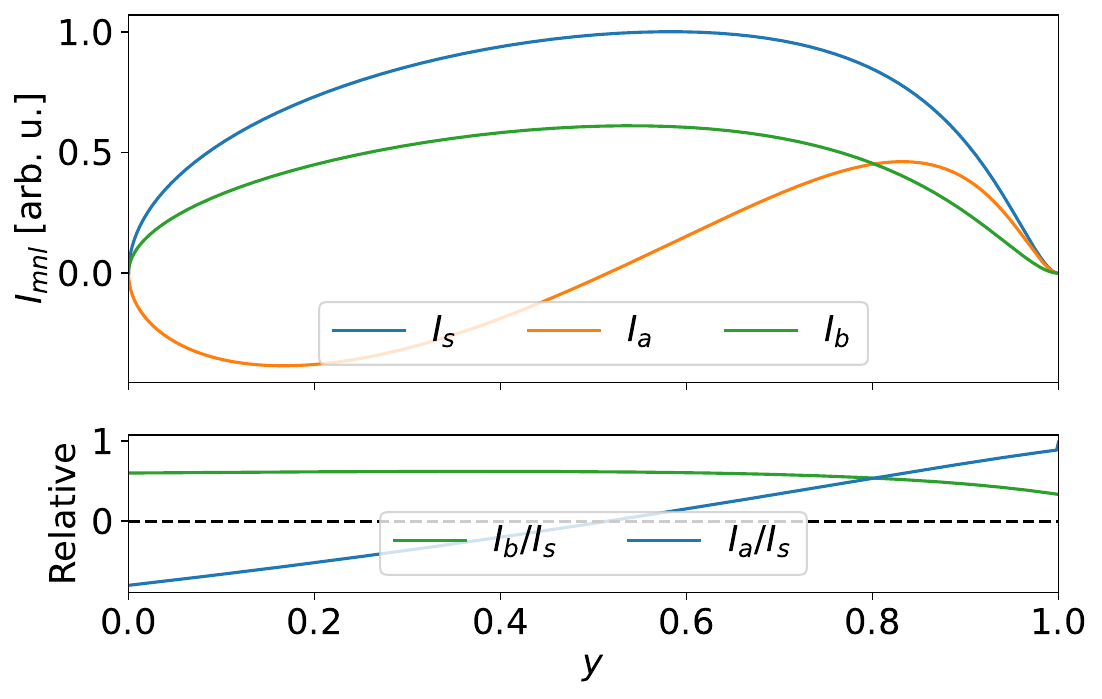}
    \caption{(Top) Normalized shapes for the elements of a tree-level recoil spectrum with $I_s = I_{110}$, $I_b = m_e I_{010}$ and $I_a = (p_f^2I_{000}-I_{002}-I_{020})/2$ as a function of $y$ where $E_f = m_f + (E_f^{max}-m_f)y$. (Bottom) Ratio of $b$ and $a$ dependent shapes relative to the scalar part, $I_s$.}
    \label{fig:I_shapes}
\end{figure}

It is instructive to look at the behaviour of the various functions appearing in Eq. (\ref{eq:recoil_spectrum_I}). Figure \ref{fig:I_shapes} shows the three main elements of the tree-level calculation for a decay with $Q=2m_e$ and $A=11$, with very shapes similar to those of Nachtmann \cite{Nachtmann1968} for the neutron \cite{Konrad2011}.

To first order, the effect of a Fierz-like term gives a very limited spectral change as discussed in Sec. \ref{sec:BSM}. The recoil spectrum is directly sensitive to $a$, on the other hand, and produces $\mathcal{O}(1)$ changes based on the lepton opening angle.

With the $\beta$ spectrum shape corrections described in Ref. as multiplicative functions $C = 1+aE_e+b/E_e+cE_e^2+d$, is it worthwhile to consider their relative effects. As with the Fierz term, $E_e^{-1}$ contributions change the recoil spectrum shape very little, and many $I_{mn0}$ end up rather similar. This conclusion can be extended to many of the spectral corrections, with spectral changes of origin $E^{1,2}_e$ largely occurring near the maximal recoil kinetic energy. A different conclusion is reached for corrections to the angular correlation. As the spectrum is directly proportional to $a$, changes due to, e.g., nuclear structure giving
\begin{equation}
    a_{\beta\nu} \to a_{\beta\nu}^0(1+\alpha_1 E_e + \alpha_2E_e^2)
\end{equation}
as discussed in Ref. \cite{Hayen2020a} give substantial corrections to the recoil shape. We will discuss this in greater detail in the following section, and show how - contrary to $\beta$ spectroscopy - many of the dominant nuclear structure uncertainties are heavily suppressed.

\section{Spectral corrections}
\label{sec:spectral_corrections}
The ability to extract Beyond Standard Model physics from the recoil spectrum requires a Standard Model prediction that is at least as accurate as the signature that is being searched for. Corrections beyond the leading-order expression generally find their origin in one of three categories: ($i$) kinematic corrections due to the three-body nature of the decay; ($ii$) higher-order corrections in the weak vertex that depend on nuclear structure; ($iii$) radiative corrections due to real and virtual photons. We will discuss each of these in turn and show how they affect the recoil spectrum and cause systematic shifts to the extracted observables of interest would they not be taken into account.

Before we do so, we recall a result from 1959 by Weinberg that is central to several of these results and is worth keeping in mind throughout. At its heart is a statement about symmetries with respect to the interchangeability of the leptons. Specifically, \textit{if a scalar correlation is measured, then the interference terms between vector and axial vector will be antisymmetric under the exchange of $l$ and $\nu$, while the `pure' terms will be symmetric}. The previous statement is generally true when the masses of the leptons can be neglected and electromagnetic interactions turned off. When no lepton spins or momenta are directly observed, however, the statement remains true when lepton masses cannot be neglected.

At tree level, the recoil kinetic energy is the scalar quantity constructed from
\begin{equation}
    T_R = \frac{|\bm{p}_e+\bm{p}_\nu|^2}{2M}
\end{equation}
and is clearly symmetric under the exchange of the leptons. We will therefore see how this plays out in the next sections.

\subsection{Kinematic corrections: Exact energies and phase space boundaries}
\label{sec:phase_space}
In much of the traditional $\beta$ decay literature the initial and final states are considered infinitely massive, and kinematic recoil corrections are typically either neglected or introduced to first order as a multiplicative factor. Following Seng's recent work and writing the phase space integral, for example, one finds
\begin{align}
    d\Gamma &= \frac{1}{512\pi^5m_i}\int d\Omega_ed\Omega_\nu \int_{m_e}^{E_0} dE_ep_e \nonumber \\
    &\times \int dE_\nu \frac{E_\nu}{m_i-E_e+p_ec} |\mathcal{M_\beta}|^2 \nonumber \\
    &\times\delta\left(E_\nu-\frac{E_0-E_e}{1-(E_e-p_ec)/m_i}\right)
    \label{eq:phase_space_integral_Seng}
\end{align}
with $c \equiv \cos \theta_{\beta\nu}$, $E_f = m_i-E_e+p_ec$ is the final state energy and $\mathcal{M}_\beta$ the tree-level $\beta$ decay matrix element. Recognizing that $|\mathcal{M}_\beta| \propto E_eE_\nu$ (see Eq. (\ref{eq:W0}), one often introduces the additional multiplicative factor
\begin{equation}
    E_\nu^2\frac{m_f}{E_f} \approx (E_0-E_e)^2\left\{1+\frac{3E_e-E_0-3p_ec}{M}\right\}
    \label{eq:recoil_approx_multiplicative}
\end{equation}
present in, e.g., the Holstein description \cite{Holstein1974}. A direct use of this formula is prone to error, however, depending on how the kinematic boundaries are defined for a recoil spectrum. Equation (\ref{eq:recoil_approx_multiplicative}) implies an additional kinematic contribution to $a$, $\delta a = -3E_e/M$, a non-zero $(\bm{p}_e\cdot \bm{p}_\nu)^2$ contribution, and an overall shift of order $E_e/M$.

Rather than continue in this fashion, we will construct the recoil spectrum directly rather than perform the two-step translation of Eqs. (\ref{eq:JTW_angular}) and (\ref{eq:dcostheta}). The available phase space in the Dalitz plot is usually interpreted through the three critical points where one of the particles has zero kinetic energy discussed in the previous section. The available phase space then lies in the connections between these points and corresponds to when $|\cos \theta_{\beta\nu}| \leq 1$ as mentioned above. In the literature, three levels of approximation can be found depending on what order the kinetic energy of the recoiling particle is taken into account. Often, one neglects the latter and one finds \cite{Gluck1993}
\begin{equation}
    E_{e}^{\mp,0}(E_f) = \frac{1}{2}\left(m_i-E_f\mp p_f+\frac{m_e^2}{m_i-E_f\mp p_f}\right)
    \label{eq:Ee_Gluck}
\end{equation}
with the reverse relationship, $E_{f}^{\mp,0}(E_e)$ obtained by replacing $e\leftrightarrow f$. Nachtmann extended this result to first order in $\Delta/M$ to find \cite{Nachtmann1968}
\begin{equation}
    E_e^{\mp,\delta} = E_e^{\mp,0}(1-\delta)+\frac{1}{2}\delta\frac{\Delta^2+m_e^2-p_f^2}{\Delta}
    \label{eq:Ee_Nachtmann}
\end{equation}
with $\delta = \Delta/2M$.

Finally, parametrizations from kaon decays have recently been used in the context of beta decay, which are exact and are reproduced here for convenience. For an electron energy $E_e$, the recoil energy is bounded by \cite{Seng2024}
\begin{subequations}
    \begin{align}
    E_f^\pm(E_e) &=  \frac{m_i}{2}(a(y)\pm b(y))\\
    a(y) = &\frac{(2-y)(1+r_f+r_e-y)}{2(1+r_e-y)},\\
    b(y) = &\frac{\sqrt{y^2-4r_e}(1+r_e-r_f-y)}{2(1+r_e-y)},
\end{align}
\label{eq:Ef_Seng}
\end{subequations}
with the dimensionless variable $y=2E_e/m_i$ evaluated in the rest frame of the parent state, and $r_{e,f} = m_{e,f}^2/m_i^2$. As before, the opposite result can be obtained by performing the substitutions $z = 2E_f/m_i\leftrightarrow y$ and $e\leftrightarrow f$.

\begin{figure}[ht]
    \centering
    \includegraphics[width=\linewidth]{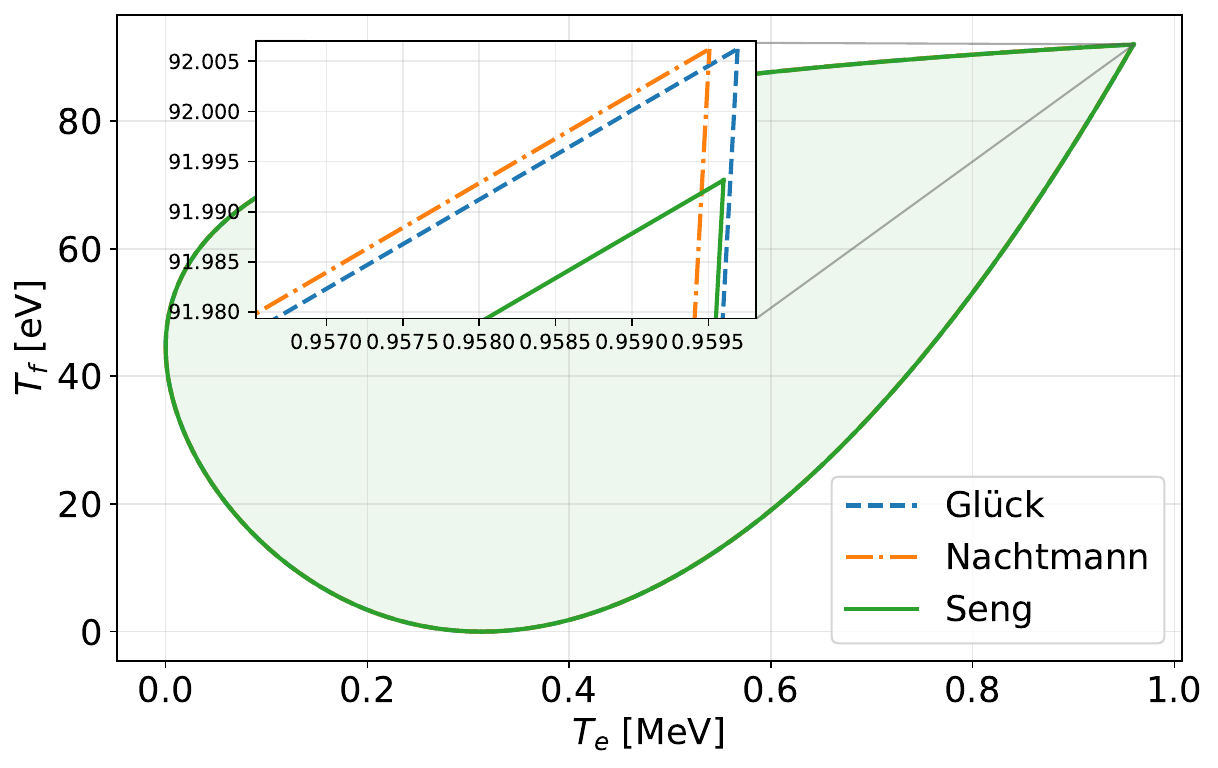}
    \caption{Comparison of different approximations of the phase space boundaries for the $^{11}$C $\beta^+$ decay. Here, Gl\"uck corresponds to a neglect of the recoil energy \cite{Gluck1993}, Eq. (\ref{eq:Ee_Gluck}), Nachtmann takes into account first-order corrections \cite{Nachtmann1968}, Eq. (\ref{eq:Ee_Nachtmann}), and Seng to the exact results \cite{Seng2024}, Eq. (\ref{eq:Ef_Seng}). }
    \label{fig:11C_recoil_boundaries}
\end{figure}

Figure \ref{fig:11C_recoil_boundaries} shows the phase space boundaries for $^{11}$C $\beta^+$ decay. As anticipated, changes are small ($\delta \approx 7\times 10^{-5}$) except for the exact kinematic endpoints. Differences are less than a keV for the positron endpoint and only 20 meV for the recoiling ion, but 

What to make of the traditional procedure of Eq. (\ref{eq:recoil_approx_multiplicative})? The correct procedure is to use the exact phase-space boundaries and the complete $E_\nu = E_0-E_e-T_f$ in the integration of Eq. (\ref{eq:phase_space_integral_Seng}), and by extension in Eq. (\ref{eq:recoil_integral}), without the additional prefactor of Eq. (\ref{eq:recoil_approx_multiplicative}). The latter would constitute a double counting, and introduce a bias in an extracted $a_{\beta\nu}$ value.

\begin{figure}
    \centering
    \includegraphics[width=\linewidth]{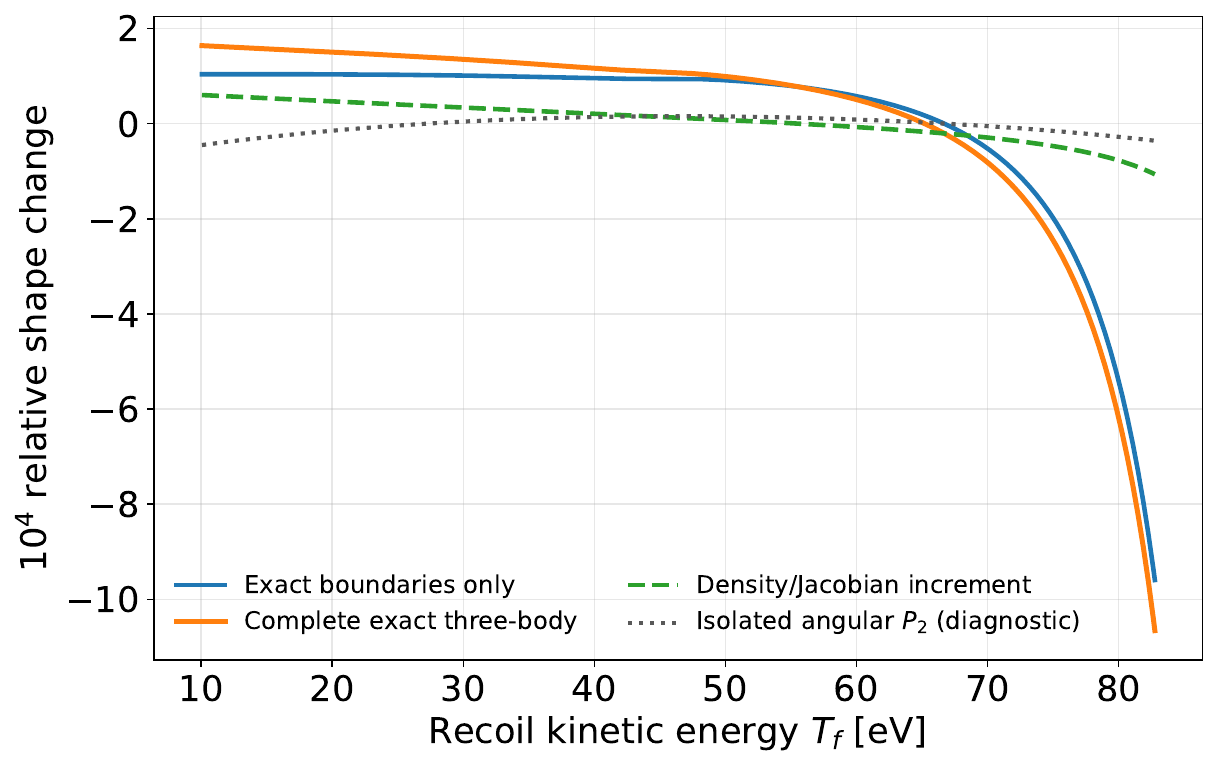}
    \caption{Relative change to the spectral shape of the recoiling nucleus following $^{11}$C $\beta^+$ decay, depending on the various approximations taken for the treatment of the phase space boundaries and potential double counting.}
    \label{fig:11C_three_body_phase_space}
\end{figure}

In fact, we may now consider what the anticipated effect of these kinematic corrections are to a truthful extraction of $a_{\beta\nu}$ from a recoil spectrum. Figure \ref{fig:11C_three_body_phase_space} shows the spectral changes for the recoil taking the relevant corrections into account for the same $^{11}$C decay. 

While most of the relevant change occurs close to the endpoint, the spectrum receives an overall slope change that directly influences an $a_{\beta\nu}$ extraction. For the latter, we perform an Asimov-style analysis and fit the spectra generated using the Gl\"uck bounds using the exact relations. The results are shown in Table 

\begin{ruledtabular}
    \begin{table}[tb]
  \centering
  \caption{Bias in the fitted beta--neutrino correlation coefficient when the
  complete exact three-body recoil spectrum, using the exact kinematic bounds
  and exact neutrino energy, is fitted with the leading Gl\"uck recoil model.
  The normalization is free and the common fit interval runs from 10 eV (or
  zero when necessary) to $0.90\,T_{\max}$.  Here
  $\delta a_{\mathrm{kin}}^{\mathrm{full}}=
  a_{\mathrm{fit}}-a_{\mathrm{SM}}$, where $a_{\mathrm{SM}}$ is obtained from
  the lifetime-derived mixing ratio as in Vanlangendonck \emph{et al.}; for
  the pure Gamow--Teller transitions, $a_{\mathrm{SM}}=-1/3$.  These values
  include the complete leading-order kinematic and phase-space effect, but not
  recoil-order form factors, radiative corrections, or detector effects.}
  \label{tab:recoil-bound-bias}
  \begin{tabular}{lccc}
    Isotope & $\Delta$ [MeV] & $\delta a_{\mathrm{kin}}^{\mathrm{full}}$
    & $\delta a_{\mathrm{kin}}^{\mathrm{full}}/a_{\mathrm{SM}}$ \\
    \midrule
    $n$ & 1.293367 & $-5.09\times10^{-3}$ & $4.76\times10^{-2}$ \\
    $^{3}\mathrm{H}$ & 0.529590 & $-1.40\times10^{-2}$ & $1.60\times10^{-1}$ \\
    $^{6}\mathrm{He}$ & 4.092999 & $-1.07\times10^{-3}$ & $3.22\times10^{-3}$ \\
    $^{18}\mathrm{F}$ & 1.144501 & $-4.15\times10^{-4}$ & $1.24\times10^{-3}$ \\
    $^{11}\mathrm{C}$ & 1.470690 & $-3.84\times10^{-4}$ & $-7.43\times10^{-4}$ \\
    $^{13}\mathrm{N}$ & 1.709471 & $-2.46\times10^{-4}$ & $-3.61\times10^{-4}$ \\
    $^{15}\mathrm{O}$ & 2.243181 & $-1.50\times10^{-4}$ & $-2.42\times10^{-4}$ \\
    $^{17}\mathrm{F}$ & 2.249471 & $-2.05\times10^{-4}$ & $-1.25\times10^{-3}$ \\
    $^{19}\mathrm{Ne}$ & 2.728501 & $-1.87\times10^{-4}$ & $-4.62\times10^{-3}$ \\
    $^{21}\mathrm{Na}$ & 3.035920 & $-7.81\times10^{-5}$ & $-1.42\times10^{-4}$ \\
    $^{23}\mathrm{Mg}$ & 3.545180 & $-4.44\times10^{-5}$ & $-6.47\times10^{-5}$ \\
    $^{25}\mathrm{Al}$ & 3.765809 & $-6.33\times10^{-5}$ & $-1.34\times10^{-4}$ \\
    $^{27}\mathrm{Si}$ & 4.301359 & $-4.39\times10^{-5}$ & $-7.78\times10^{-5}$ \\
    $^{29}\mathrm{P}$ & 4.431231 & $-2.64\times10^{-5}$ & $-3.77\times10^{-5}$ \\
    $^{31}\mathrm{S}$ & 4.887011 & $-2.25\times10^{-5}$ & $-3.18\times10^{-5}$ \\
    $^{33}\mathrm{Cl}$ & 5.071521 & $-9.64\times10^{-6}$ & $-1.09\times10^{-5}$ \\
    $^{35}\mathrm{Ar}$ & 5.455241 & $-7.36\times10^{-6}$ & $-8.17\times10^{-6}$ \\
    $^{37}\mathrm{K}$ & 5.636481 & $-2.16\times10^{-5}$ & $-3.24\times10^{-5}$ \\
    $^{39}\mathrm{Ca}$ & 6.013491 & $-2.76\times10^{-5}$ & $-4.63\times10^{-5}$
  \end{tabular}
\end{table}
\end{ruledtabular}

The overall effect is to reduce the extracted value of $a_{\beta\nu}$. As anticipated, the shift is inversely proportional the mass and absolute shifts are never larger than a few per-mille. For the neutron this corresponds to a relative change of almost $5\%$, however, due to the substantial cancellation in the tree-level result and its low mass. The triton, with its maximal recoil energy of 3.4 eV, is clearly a special case and included for exhaustiveness. For $^{19}$Ne, where such a cancellation is even more extreme, the relative change to $a$ is only at the per-mille scale. For the well-studied case of $^6$He, the change is below the current level of sensitivity but nn-negligible. Note that the results obtained depend strongly on the chosen fit window, however, as can be seen from Fig. \ref{fig:11C_three_body_phase_space}.

\subsection{Center of mass Fermi function}

The traditional Fermi function is defined through the interaction of the final state $\beta$ particle with a stationary Coulomb potential. As the recoiling nucleus is moving away from the $\beta$ particle, however, a small correction is introduced by looking at the relative velocity as a function of $\beta$ and recoil energies. This was originally introduced by Wilkinson \cite{Wilkinson1982} for the $\beta$ spectrum in the non-relativistic approximation.

We may directly generalize the result, writing
\begin{equation}
    \beta_\mathrm{rel} = \sqrt{1-\left(\frac{m_em_f}{p_e\cdot p_f}\right)^2}
\end{equation}
Taking first the simple Fermi function approximation, the correction is simply
\begin{align}
    Q(Z, E_e, E_f) &= 1+\alpha Z \pi \left(\beta_\mathrm{rel}^{-1}-\beta^{-1}_e\right) \nonumber \\
    & = 1+\alpha Z \pi \frac{m_e^2}{m_fp_e^2}p_f\cos\theta_{ef}
\end{align}
which may easily be generalized as
\begin{equation}
    Q(E, E_e, E_f) = 1-(1-\beta_e^2)\frac{p_f}{m_f}\cos\theta_{ef}\frac{\partial \ln F}{\partial \beta_e}
\end{equation}

The weighting with the $\beta$-$\nu$ angular correlation results in the clean $a_{\beta\nu}$ dependence in Wilkinson's $Q(Z, E_e)$ result. The latter is visible at extremely low energies in the $\beta$ spectrum. We may then derive the recoil spectrum correction to be
\begin{align}
 Q_{\mathrm{rec}}(p_f)&=1+
 \frac{\pi\alpha Z m_e^2}{2m_f} \nonumber \\
 &\times\frac{\displaystyle
  \int_{E_-}^{E_+}\dd\ee\,W(\ee,E_f)
  \frac{\enu^2-\pe^2-p_f^2}{\pe^3}}
 {\displaystyle
  \int_{E_-}^{E_+}\dd\ee\,W(\ee,E_f)}
 \label{eq:qrec-main}
\end{align}

\begin{figure}[ht]
    \centering
    \includegraphics[width=\linewidth]{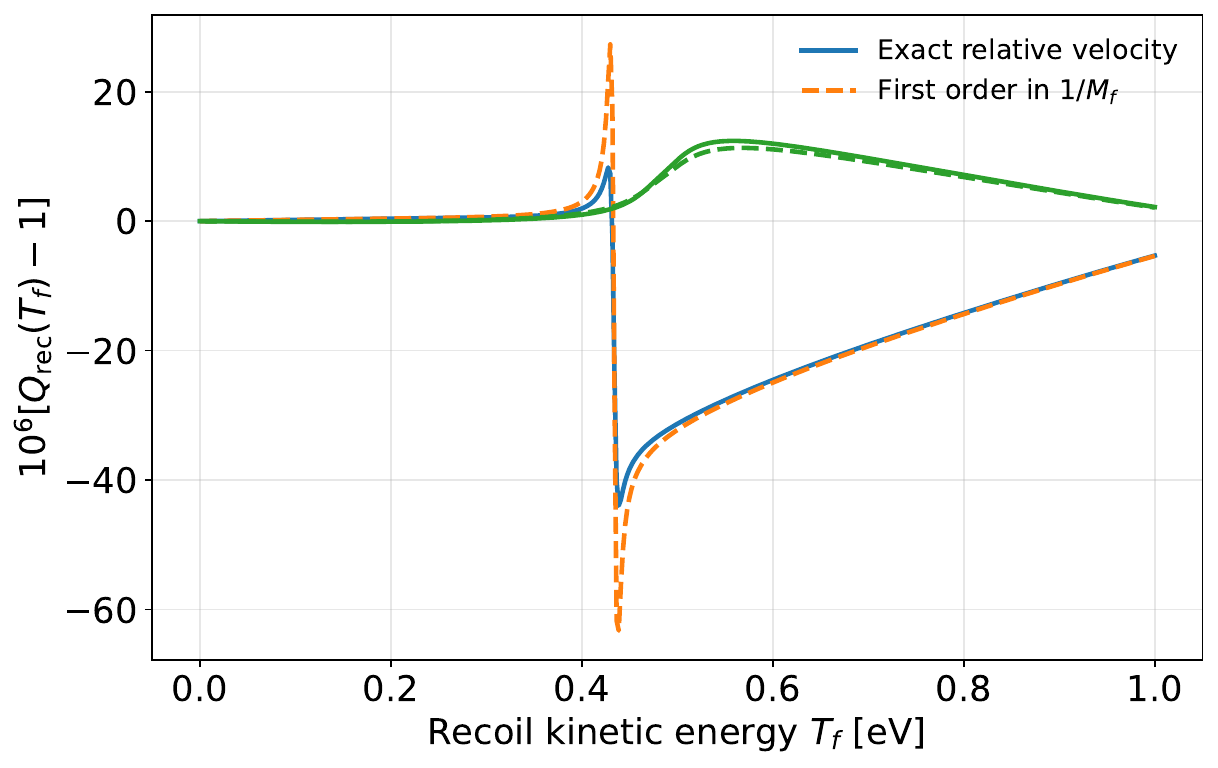}
    \caption{Relative recoil spectral correction when the Fermi function is evaluated in the co-moving frame of the recoiling nucleus for neutron and $^{11}$C decays. }
    \label{fig:n_recoil_Q}
\end{figure}

Figure \ref{fig:n_recoil_Q} shows the effect on the neutron spectrum, where its effect is greatest, and for $^{11}$C $\beta^+$ decay with a more modest effect. The former displays a logarithmic divergence when the electron and proton become co-moving and the Coulomb interaction diverges, when $\beta_e \approx 8\times 10^{-4}$. For the $\beta^+$ decay of $^{11}$C this is never the case, and the entire function behaves cleanly over the entire interval. The logarithmic divergence for $\beta^-$ decays in integrable, however, and poses no problem for experimental analyses. 

While the logarithmic divergence in a $\beta^-$ decay can maybe cause observable shifts in some future scenario, the induced absolute shift in an extraction of $a_{\beta\nu}$ is only of order $10^{-5}$ for the neutron, and can be ignored for the other isotopes. 

\subsection{Spectral corrections}

The machinery developed for the accurate prediction of allowed $\beta$ spectrum shapes \cite{Hayen2018, Hayen2019a} may be transferred directly to that of the recoil shape with only a few exceptions. Many of these corrections depend only on the electron energy and are electrostatic in nature, such as the nuclear finite size, screening and atomic exchange corrections. Corrections pertaining to nuclear structure, on the other hand, are discussed explicitly in the following section. As many of these corrections are dominant at low $\beta$ particle energies, these show up around the $B$ critical point in Fig. \ref{fig:Dalitz_pedagogical}.

\begin{figure}[ht]
    \centering
    \includegraphics[width=\linewidth]{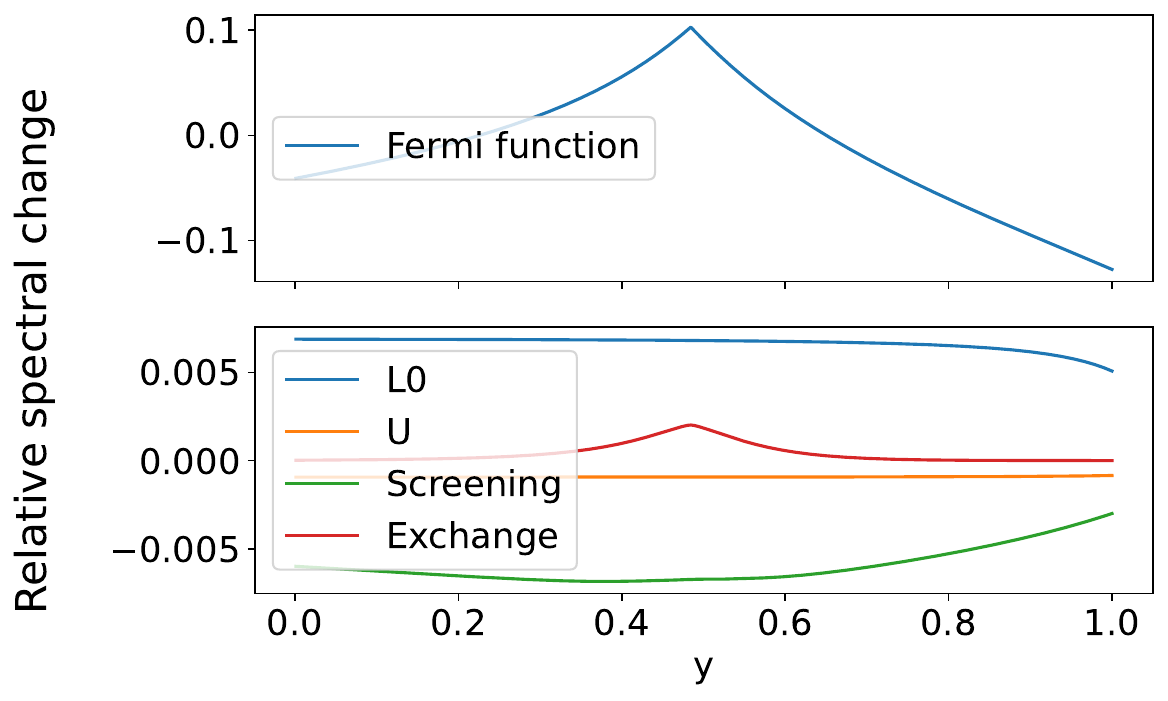}
    \caption{Relative corrections to the recoil spectrum using the correction functions to the $\beta$ spectrum shape for a $Z=20, A=40, E_0 = 1.5$ MeV decay. The top panel shows the effect of the Fermi function for a normalized spectrum, whereas the bottom shows absolute changes. }
    \label{fig:spectral_corrections}
\end{figure}

Figure \ref{fig:spectral_corrections} shows the effects of several typical correction factors to the $\beta$ spectrum shape for a fictitious $Z=20, A=40, E_0 = 1.5$ MeV decay. Naming and more details may be found in Refs. \cite{Hayen2018, Hayen2019a}. Many of the corrections concern final state interactions between the $\beta$ particle and recoil, which are dominant at low $E_e$. This is reflected in the maximal shape deviations occurring at the recoil energy when $E_e\to 0$. For large $Z$, the Fermi function is significant and non-constant throughout the entire $\beta$ spectrum. As a result, its influence on the recoil spectrum is similarly spread out.

Overall, because the recoil spectrum contains the integral over all possible $\beta$ particle energies consistent with $|\cos\theta_{\beta\nu}| \leq 1$, the magnitude of spectral change is typically at least an order of magnitude lower than than in the $\beta$ spectrum. While this significantly eases the theoretical control required for obtaining a high precision recoil spectrum, the sensitivity of, e.g., an $a_{\beta\nu}$ extraction to small spectral changes means that the figure of merit is case-dependent and harder to evaluate.

\subsection{Recoil-order corrections}

In the previous subsection we have focused on the electrostatic corrections that directly change the $\beta$ spectrum shape. Of more direct concern are the recoil-order corrections to the weak vertex. Generalizations of the hadronic current have been performed in various different formalisms (see Refs. \cite{Hayen2018, Hayen2026} for an overview). For conceptual clarity we use the framework by Holstein \cite{Holstein1974} to demonstrate the main results but the numerical changes are obtained from calculations within the Behrens-B\"uhring formalism \cite{Behrens1971, Behrens1982}.

Following Ref. \cite{Gluck1998}, we may generalize the Dalitz distribution to take into account the leading recoil-order matrix elements as follows
\begin{subequations}
    \begin{align}
    W_0 &= \frac{G_F^2V_{ud}^2}{4\pi^3}\left[(f_1 +f_2)E_eE_\nu + f_2m_i(E_f-E_f^{max})\right],\\
    f_1 &= a^2+c^2-\frac{2}{3}\frac{E_0}{M}(c^2\pm cb \pm cd) \nonumber \\
    &+ \frac{2}{3}\frac{E_e}{M}(3a^2+5c^2\pm 2cb) \nonumber \\
    &-\frac{1}{3}\frac{m_e^2}{ME_e}(2c^2\pm2cb\pm2cd)\\
    f_2 &= a^2-\frac{1}{3}c^2+\frac{2}{3}\frac{E_0}{M}(c^2\pm cb\pm cd)\nonumber \\
    &-\frac{4}{3}\frac{E_e}{M}(3c^2\pm cb)
\end{align}
\label{eq:W0_f1_f2}
\end{subequations}
Here $a(c)$ is the usual Fermi (Gamow-Teller) form factor, $b$ is called weak magnetism and $d$ is the induced tensor. Unless particular cancellations occur, one has $b/Ac \sim 5$ for mass number $A$ \cite{Severijns2023}, such that generally $b\gg c, a$. Generically $d$ is of similar magnitude, but is constrained to zero within isospin multiplets (such as mirror decays) \cite{Holstein1974} but unconstrained for generic Gamow-Teller transitions \cite{King2022, Glick-Magid2022}. 

As such, weak magnetism is usually the dominant contribution to the $\beta$ spectrum shape. More importantly for this topic, however, are Weinberg's findings mentioned at the start of this section. The weak magnetism form factor, $b$, is of vector origin whereas its influence in Eqs. (\ref{eq:W0_f1_f2}) is through multiplication of the axial vector form factor $c$\footnote{This can easily be understood by noting that both appear at leading order in the spatial part of the weak hadronic current, unlike the timelike vector form factor $a$.}. 

In the absence of Coulomb interactions, therefore, the effect of the weak magnetism form factor on the recoil spectrum is \textit{exactly equal to zero}. We may note already that in summing $f_1+f_2$ the main $b$ and $d$ dependent terms cancel, leaving only the $E_e^{-1}$ term. Explicit calculation shows that for each $E_f$ all $b$ dependent terms integrate to zero as expected. This also allows us to write down the identity
\begin{equation}
    -\frac{m_e^2}{M}I_{010}+(E_0I_{000}-2I_{100})(E_f-E_f^{max}) = 0
\end{equation}
for all energies $E_f$.

Turning on the Coulomb interaction, we may derive the reduced influence on the recoil spectrum as
\begin{align}
    &\delta_{wm}(E_f) = \frac{\alpha Z \pi}{3}cb[\Omega_{\alpha b}(E_e^+(E_f))-\Omega_{\alpha b}(E_e^-(E_f))], \\
    &\Omega_{\alpha b}(E_e) = \frac{m_e^2}{M}\left[p_e(E_e-E_0)+m_e^2\ln\left(\frac{1+\beta}{1-\beta}\right)\right] \nonumber \\
    &- (E_f-E_f^{max})\left[p_e(E_e-2E_0)+2m_e^2\ln\left(\frac{1+\beta}{1-\beta}\right)\right]
\end{align}
generalizing the result for the neutron \cite{Gluck1993}. We may note that the sign is unchanged regardless of $\beta^\pm$.

The influence of the induced tensor form factor, $d$, survives even without Coulomb interactions as part of the axial current generalization. Even so, one may notice that only the $E_e^{-1}$ and $E_0$ contributions survive. The recoil spectrum is largely free of recoil-order corrections, therefore, and as a consequence the uncertainties that play a significant role in $\beta$ spectroscopy are largely negligible when performing recoil spectroscopy.

Nevertheless, neglecting to account for these additional terms causes systematic shifts in an extracting of $a_{\beta\nu}$ from experimental data, while the uncertainty on the correction produces a precision ceiling on any potential Beyond Standard Model physics extracted from it. We therefore calculate the shifts introduced by the various corrections on $a_{\beta\nu}$ for all mirror isotopes with $A<40$, shown in Table \ref{tab:weak-magnetism-pseudoscalar-bias}.

\begin{ruledtabular}
\begin{table*}[tb]
  \centering
  \footnotesize
  \setlength{\tabcolsep}{4pt}
  \caption{Isolated Asimov shifts in the fitted beta--neutrino
  correlation coefficient from weak magnetism and the induced
  pseudoscalar form factor.  For the nuclear mirror transitions,
  $c$ and $b$ (with parenthetical one-standard-deviation
  uncertainties) are from Table XIV of Ref. \cite{Severijns2023},
  while $h/A^2c$ is from Table XIX and is shown to two significant
  digits because no uncertainty is quoted.  The neutron is not
  tabulated in those tables and is retained only as a benchmark.
  The normalization is free and the fit interval runs from 10 eV
  (or zero when necessary) to $0.90\,T_{\max}$.  The shifts change
  one form factor at a time relative to $(b,d,h)=(0,0,0)$.}
  \label{tab:weak-magnetism-pseudoscalar-bias}
  \begin{tabular}{lrrrrrr}
    Name & $c$ & $b$ & $h/A^2c$ & $a_0$ & $\delta a_b$ & $\delta a_h$ \\
    \midrule
    $n$ & $2.2$ & $8.2$ & -- & $-0.10688$ & $-3.9\times10^{-5}$ & -- \\
    $^{3}\mathrm{H}$ & $-2.1053(12)$ & $-26.53394540(16)$ & $1.8\times10^{2}$ & $-0.087887$ & $-2.3\times10^{-4}$ & $2.8\times10^{-5}$ \\
    $^{11}\mathrm{C}$ & $-0.75442(28)$ & $-51.871(14)$ & $1.1\times10^{2}$ & $0.51638$ & $3.5\times10^{-4}$ & $-6.2\times10^{-5}$ \\
    $^{13}\mathrm{N}$ & $-0.55962(34)$ & $-23.0708(90)$ & $-39$ & $0.68202$ & $1.2\times10^{-4}$ & $2\times10^{-5}$ \\
    $^{15}\mathrm{O}$ & $0.63023(46)$ & $26.0506(31)$ & $-82$ & $0.62096$ & $8.5\times10^{-5}$ & $7.6\times10^{-5}$ \\
    $^{17}\mathrm{F}$ & $1.29555(65)$ & $133.0616(61)$ & $1.3\times10^{2}$ & $0.16447$ & $1.2\times10^{-4}$ & $-2.7\times10^{-4}$ \\
    $^{19}\mathrm{Ne}$ & $-1.60203(65)$ & $-148.5605(26)$ & $1.8\times10^{2}$ & $0.040516$ & $-7.5\times10^{-6}$ & $-5.9\times10^{-4}$ \\
    $^{21}\mathrm{Na}$ & $0.71245(39)$ & $82.6366(27)$ & $1.6\times10^{2}$ & $0.55108$ & $1.2\times10^{-4}$ & $-3.2\times10^{-4}$ \\
    $^{23}\mathrm{Mg}$ & $-0.55413(47)$ & $-81.7768(63)$ & $1.3\times10^{2}$ & $0.68677$ & $1\times10^{-4}$ & $-2.4\times10^{-4}$ \\
    $^{25}\mathrm{Al}$ & $0.80844(42)$ & $133.140(36)$ & $94$ & $0.473$ & $9.2\times10^{-5}$ & $-3.1\times10^{-4}$ \\
    $^{27}\mathrm{Si}$ & $-0.69659(30)$ & $-143.9792(77)$ & $57$ & $0.56439$ & $9.2\times10^{-5}$ & $-1.9\times10^{-4}$ \\
    $^{29}\mathrm{P}$ & $0.53798(47)$ & $89.913(11)$ & $45$ & $0.70072$ & $6.4\times10^{-5}$ & $-1.1\times10^{-4}$ \\
    $^{31}\mathrm{S}$ & $-0.52939(33)$ & $-86.9584(46)$ & $39$ & $0.70813$ & $5\times10^{-5}$ & $-1.1\times10^{-4}$ \\
    $^{33}\mathrm{Cl}$ & $-0.31416(28)$ & $4.728(21)$ & $-1.9\times10^{2}$ & $0.88023$ & $-2.4\times10^{-6}$ & $2.5\times10^{-4}$ \\
    $^{35}\mathrm{Ar}$ & $0.28199(17)$ & $-8.5704(90)$ & $-1.9\times10^{2}$ & $0.90179$ & $-3.5\times10^{-6}$ & $2.2\times10^{-4}$ \\
    $^{37}\mathrm{K}$ & $-0.57789(39)$ & $-44.99(24)$ & $-51$ & $0.6662$ & $1.5\times10^{-5}$ & $2.2\times10^{-4}$ \\
    $^{39}\mathrm{Ca}$ & $0.66061(50)$ & $31.7295(61)$ & $-53$ & $0.59491$ & $6.8\times10^{-6}$ & $3.1\times10^{-4}$
  \end{tabular}
\end{table*}
\end{ruledtabular}

As anticipated, shifts due to weak magnetism on the extraction of $a_{\beta\nu}$ are very small and beyond current experimental precision. A similar conclusion is reached for the $h(q^2)$ form factor. Contrary to regular $\beta$ spectroscopy then, the effects of recoil-order matrix elements are subdominant and below the level of current experimental precision.

\begin{ruledtabular}
\begin{table*}[tb]
  \centering
  \small
  \setlength{\tabcolsep}{6pt}
  \caption{Corresponding isolated Asimov study for the pure
  Gamow--Teller transitions.  The experimental $c$ and $b$ values
  are from Ref. \cite{Severijns2023}, with signs fixed by the
  convention used in the recoil calculation.  For $^{6}\mathrm{He}$,
  $d=7.9$ is adopted from Ref. \cite{King2022}; The $^{18}$F value was calculated using the nuclear shell model with the USDB interaction using KSHELL \cite{Shimizu2013}.}
  \label{tab:pure-gt-matrix-element-bias}
  \begin{tabular}{lrrrrrr}
    Name & $c$ & $b$ & $d$ & $a_0$ & $\delta a_b$ & $\delta a_d$ \\
    \midrule
    $^{6}\mathrm{He}$ & $-2.7802(20)$ & $-68.2(7)$ & $7.9$ & $-1/3$ & $1.6\times10^{-4}$ & $9.2\times10^{-4}$ \\
    $^{18}\mathrm{F}$ & $-1.2778(22)$ & $-133(13)$ & 13.8 & $-1/3$ & $8.3\times10^{-5}$ & $3.3\times10^{-4}$
  \end{tabular}
\end{table*}
\end{ruledtabular}

A similar conclusion is reached for the pure Gamow-Teller decays shown in Table \ref{tab:pure-gt-matrix-element-bias}. The dominant effect comes from the induced tensor form factor,

\subsection{Radiative corrections}

The effects of radiative corrections on recoil spectra have recently been highlighted once more following a substantial body of work in the late 1980's and 1990's. The salient feature is that the radiative corrections derived analytically assuming the detection of the (anti)neutrino are not appropriate when instead the recoiling particle is measured, and solutions are generally obtained numerically only. The underlying reason is simple to understand, as the emission of a real photon turns the transition into a 4-body decay and the recoil energy is instead
\begin{equation}
    T_R^\gamma = \frac{|\bm{p_e}+\bm{p}_\nu+\bm{k}|^2}{2M}
\end{equation}
for a photon with momentum $\bm{k}$. This has two immediate consequences: ($i$) the region of the Dalitz distribution below the $E_f^-$ when $E_e < E_e^m$ curve is populated solely by events where photons have a non-negligible energy (\textit{hard} photons); ($ii$) the remaining Dalitz surface is modified both by hard and virtual photons. The result is a depopulation of $\beta$ particles near the endpoint, resulting in an analogous effect in the recoil spectrum.

As with the well-known radiative corrections to the $\beta$ spectrum shape, the total effect increases with $\Delta$. The latter can also be understood intuitively by the fact that the additional 4-body phase volume increases strongly with $\Delta$. For realistic geometries and nontrivial interactions of the real photon one must instead resort to Monte Carlo event generation and propagation through the experiment. On the other hand, in a simplified setting the effects can be calculated semi-analytically following the description by Gluck 1993, or more recently by Seng 2024. The latter is appropriate, for example, when the photon is not detected and the electron signal can be cleanly separated (or not observed at all).

We calculate the radiative correction to the Dalitz distribution according to Ref. \cite{Gluck1993}. The latter separates the total $\mathcal{O}(\alpha)$ calculation into a \textit{virtual-soft} and \textit{hard} contribution depending on whether the photon energy is smaller or larger than some threshold $\omega$, respectively. The former is infrared-convergent as anticipated (both virtual and soft soft photons contributions contain separate IR divergences) but depends explicitly on $\omega$. This dependence is canceled by the hard photon emission rate in integrating from $\omega$ to $k^{max}$, so that the total result is independent of $\omega$. As mentioned in Ref. \cite{Gluck1993}, however, $\omega$ can not be chosen as a constant throughout the entire Dalitz distribution and we take
\begin{equation}
    \omega = C_S E_\nu^0
\end{equation}
with $C_S\ll 1$ and $E_\nu^0 = (E_0-E_e)$ inspired by Ref. \cite{Gluck1997}. For typical calculations we choose $C_S = 10^{-5}$.

\begin{figure}
    \centering
    \includegraphics[width=\linewidth]{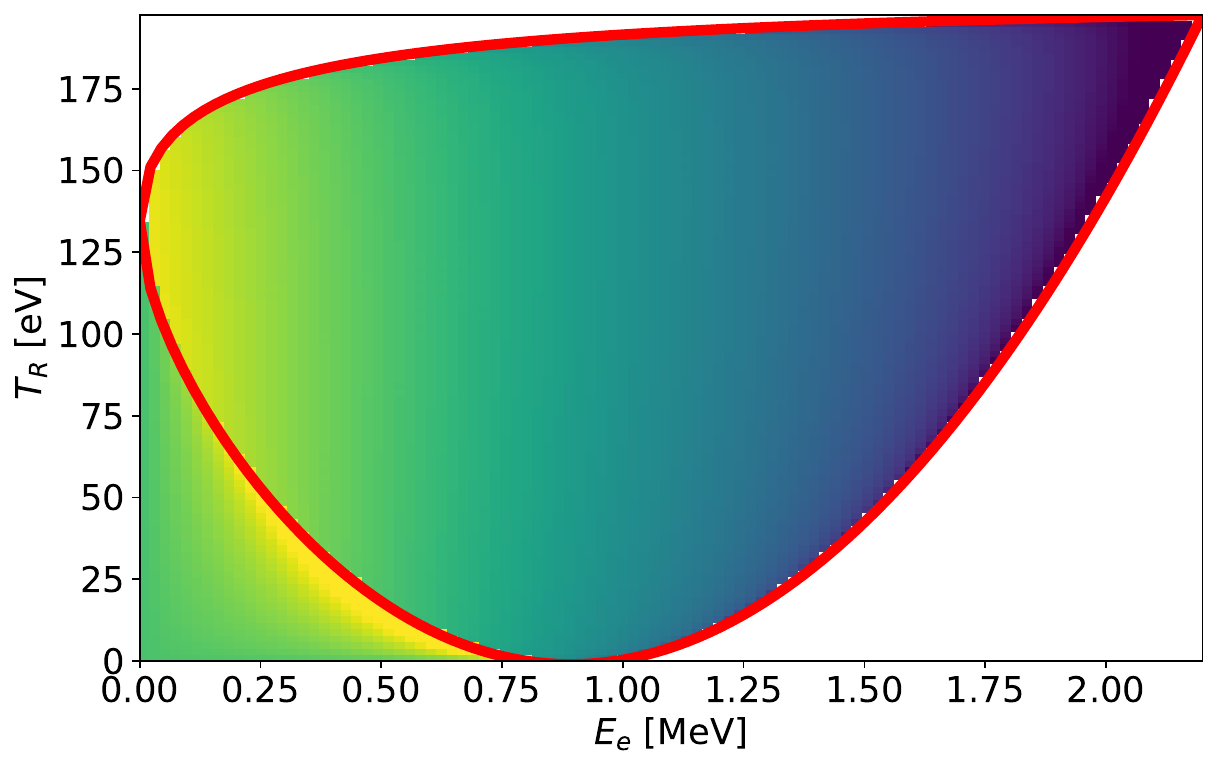}
    \caption{Relative change to the total Dalitz distribution due to radiative corrections for $^{19}$Ne $\beta^+$ decay. In the `OUT' region, the radiative amplitude is plotted relative to the mean tree-level amplitude for the same recoil energy.}
    \label{fig:RC_relative_Dalitz}
\end{figure}

Figure \ref{fig:RC_relative_Dalitz} shows the relative effect of the radiative corrections to the Dalitz distribution for $^{19}$Ne. To represent the relative contribution in the `out' region, we divide the real bremsstrahlung emission rate by the average tree-level rate at the same recoil energy. This is not a unique choice, and is purely to indicate the relative importance of the out region to the recoil spectrum. Since the phase space volume in the out region grows with $\Delta$, this is an isotope-dependent contribution. At the interface between `IN' and `OUT' regions the disappearing photon energy approaches the infrared divergence. The radiative amplitude in the `OUT' region is completely divergence-free, however, as the photon energy is always larger than a given threshold.

\begin{figure}
    \centering
    \includegraphics[width=\linewidth]{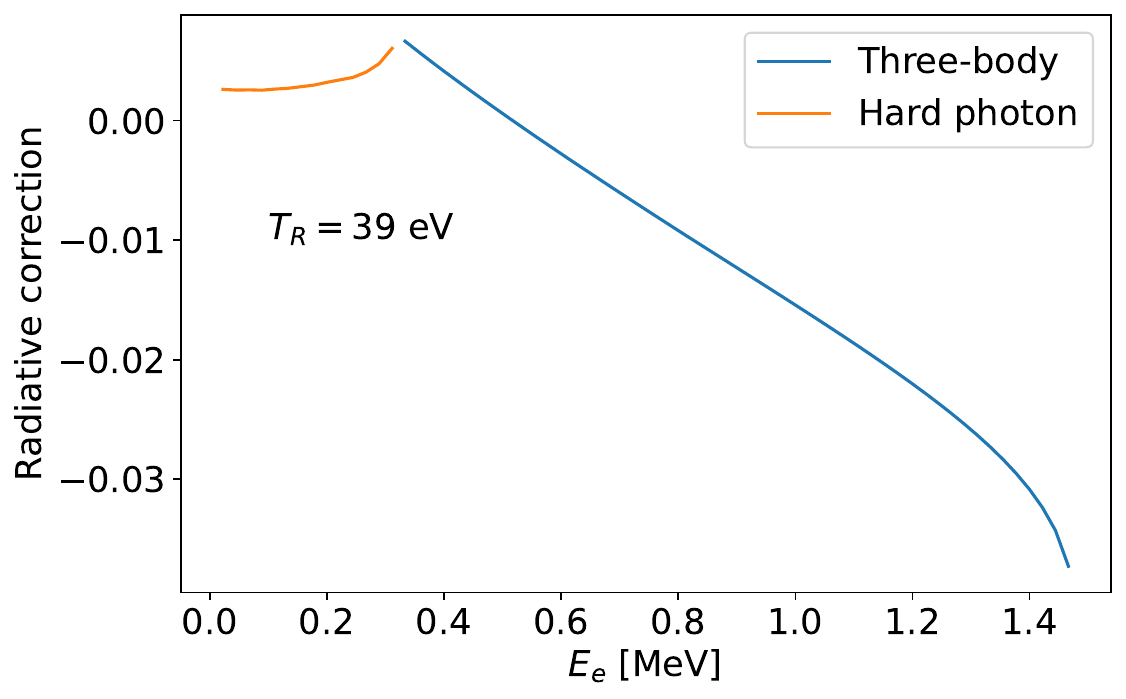}
    \caption{Radiative correction for a slice of constant recoil kinetic energy. The contribution is separated into the real and virtual photon contributions.}
    \label{fig:RC_constant_TR}
\end{figure}

We may use the results to obtain the corrections to either recoil or $\beta$ spectrum shape assuming all energies of the other particle are accepted in an experimental analysis. Doing so is a simply integration, and for the mirror isotopes we obtain the results previously published in Ref. \cite{Vanlangendonck2022}. In many experiments, however, the recoiling nucleus is detected in coincidence with the electron. In time-of-flight measurements, such as in trap experiments or Nab, the electron serves as the $t_0$ trigger, and its detection is often dependent on thresholds or detection efficiencies. Observing the radiative corrections for a slice of constant recoil energy as in Fig. \ref{fig:RC_constant_TR}, we see that near the edges of the kinematically allowed region one encounters divergences due to the shrinking of the photon phase space. These influence the averaged radiative correction, such that thresholds can play a part in the recoil radiative corrections by cutting out some of these regions. As mentioned before, however, for complex geometries and possible photon interactions or detection, a Monte Carlo generator should be used instead.

Finally, we now look at the relative influence of these radiative corrections on the determination of the $\beta$-$\nu$ angular correlation, $a_{\beta\nu}$, from a measured recoil spectrum. This has been done previous for $^6$He and $^{32}$Ar \cite{Gluck1998}, where corrections were shown to be relevant the level of precision quoted. We perform this analysis for all mirror nuclei below $A<40$ as before with a number of notable transitions. The results are shown in Table \ref{tab:recoil-radiative-bias}.

\begin{ruledtabular}
    \begin{table}[ht]
        \centering
        \caption{Bias in the fitted beta--neutrino correlation coefficient when the
  recoil-type order-$\alpha$ radiatively corrected spectrum is fitted with
  the corresponding three-body spectrum.  The normalization is free and the
  common fit interval runs from 10 eV (or zero when necessary) to
  $0.90\,T_{\max}$.  Here
  $\delta a=a_{\mathrm{fit}}-a_{\mathrm{true}}$.}
  \label{tab:recoil-radiative-bias}
  \begin{tabular}{lc}
    Isotope & $\delta a_{\mathrm{RC}}$ \\
    \midrule
    $n$ & $-2.9\times10^{-3}$ \\
    $^{3}\mathrm{H}$ & $-6.1\times10^{-3}$ \\
    $^{6}\mathrm{He}$ & $-1.9\times10^{-3}$ \\
    $^{18}\mathrm{F}$ & $-4.1\times10^{-3}$ \\
    $^{11}\mathrm{C}$ & $-1.9\times10^{-3}$ \\
    $^{13}\mathrm{N}$ & $-1.4\times10^{-3}$ \\
    $^{15}\mathrm{O}$ & $-1.2\times10^{-3}$ \\
    $^{17}\mathrm{F}$ & $-1.7\times10^{-3}$ \\
    $^{19}\mathrm{Ne}$ & $-1.7\times10^{-3}$ \\
    $^{21}\mathrm{Na}$ & $-1.1\times10^{-3}$ \\
    $^{23}\mathrm{Mg}$ & $-9.5\times10^{-4}$ \\
    $^{25}\mathrm{Al}$ & $-1.1\times10^{-3}$ \\
    $^{27}\mathrm{Si}$ & $-1.0\times10^{-3}$ \\
    $^{29}\mathrm{P}$ & $-9.7\times10^{-4}$ \\
    $^{31}\mathrm{S}$ & $-9.6\times10^{-4}$ \\
    $^{33}\mathrm{Cl}$ & $-8.9\times10^{-4}$ \\
    $^{35}\mathrm{Ar}$ & $-8.8\times10^{-4}$ \\
    $^{37}\mathrm{K}$ & $-9.8\times10^{-4}$ \\
    $^{39}\mathrm{Ca}$ & $-1.1\times10^{-3}$
    \end{tabular}
    \end{table}
\end{ruledtabular}

\section{Traditional Beyond Standard Model recoil channels}
\label{sec:BSM}

Besides initial experiments during the early days of radioactivity and nuclear $\beta$ decay, recoil spectroscopy has mostly been limited to time of flight measurements with the advent of ion and atom traps. Recently, superconducting sensors have reached the energy sensitivity required to measured the eV-scale recoils. In particular, significant success has been obtained with Superconducting Tunnel Junctions in the detection of the $^7$Li recoils following $^7$Be electron capture \cite{Fretwell2020, Friedrich2021, Kim2024, Smolsky2024}. This has opened up new avenues for research, and the recoil spectrum as a laboratory for Beyond Standard Model physics is relatively new besides those variables also looked for in typical $\beta$ spectroscopy. Nevertheless, direct energy spectroscopy opens up a key variable in these more mature searches. Here, we will specifically look at three of these: ($i$) exotic scalar and tensor current searches, typically through the Fierz interference; ($ii$) $V_{ud}$ extraction and CKM unitarity tests; ($iii$) sterile neutrino searches. This is limited by design, and we defer discussions of electron capture and more exotic signatures to future work.

\subsection{Exotic tensor interactions}
Natural extensions to the Standard Model can be obtained by enlarging the Lorentz structure of the electroweak interaction. Scalar or tensor (chiral) interactions are naturally created at tree or loop level in generic extensions, such that their search presents a wide window. If such interactions occur naturally at a high scale, an effective field theory (EFT) is a natural description for its effects at the experimental scale, denoted $\mu$\footnote{This is technically the renormalization scale in, e.g., the $\overline{MS}$ scheme.}. For $\mu \sim 2$ GeV, one may write an effective Lagrangian for the low-energy EFT (LEFT) using effective quark couplings
\cite{Holstein2014a, Cirigliano2013a},
\begin{align}
&\mathcal{L}_\text{eff} = - \frac{G_F\tilde{V}_{ud}}{\sqrt{2}}\biggl\{\bar{e}\gamma_\mu\nu_L \cdot \bar{u} \gamma^\mu[1-(1-2\epsilon_R)\gamma^5]d \\
&+\epsilon_S\, \bar{e}\nu_L \cdot \bar{u}d+ \epsilon_T\, \bar{e}\sigma_{\mu\nu}\nu_L \cdot \bar{u} \sigma^{\mu\nu}(1-\gamma^5)d +\ldots\biggr\}+ \text{h.c.}, \nonumber
\end{align}
where $G_F \approx 10^{-5}$~GeV$^{-2}$ is Fermi's constant, $\tilde{V}_{ud} = V_{ud}(1+\epsilon_L+\epsilon_R)$ the \textit{up-down} quark mixing matrix element, modified by exotic interactions denoted $\epsilon_X$. The latter are generated by dimension-6 operators and are of order
\begin{equation}
    \epsilon_X \sim \frac{v^2}{\Lambda_X^2}
\end{equation}
where $v = (2^{1/4}\sqrt{G_F})^{-1} \approx 246$ GeV is the Brout-Englert-Higgs vacuum expectation value, and $\Lambda_X$ the natural scale for the new physics. Constraints on $\epsilon_X$ at the per-mille level or below probes BSM at the (tens) of TeV scale.

In $\beta$ decay, the relevant scale is at the nuclear level, and relevant degrees of freedom are individual nucleons. The relevant EFT was written already in 1956 by Lee and Yang \cite{Lee1956}, and a matching to LEFT and has been performed in the last two decades allowing one to directly compare $\beta$ decay observables to those obtained at colliders \cite{Falkowski2017}. Allowing for exotic scalar and tensor currents as proposed in Ref. \cite{Lee1956} largely provides quadratic sensitivity in all but a few $\beta$ decay observables. The one that is most easily accessible is the Fierz interference term, which is linear in these exotic couplings and can in the modern language be written as
\begin{align}
    b_F = \pm 2 \gamma \frac{1}{1+\rho^2}&\mathrm{Re}\left\{\frac{g_S\epsilon_S}{g_V(1+\epsilon_L+\epsilon_R^\prime)} \right.\nonumber \\
    &\left.+ \rho^2 \frac{4g_T\epsilon_T}{-g_A(1+\epsilon_L - \epsilon_R^\prime)} \right\},
    \label{eq:bF}
\end{align}
Here, $\rho = g_AM_{GT}/g_VM_F$ is the Gamow-Teller to Fermi mixing ratio for an allowed decay, and $\gamma = \sqrt{1-(\alpha Z)^2}$ with $\alpha$ the fine-structure constant and $Z$ the final state proton number. The additional couplings $g_i$ are the nucleon isovector charges, defined such that $\bar{d}\Gamma_iu = g_i\bar{n}\Gamma_ip$. Lattice QCD finds the following values for them $g_S = 1.02(10)$ and $g_T = 0.989(34)$ \cite{Aoki2021}.

The Fierz interference term adds an additional energy-dependent term in the differential decay as shown in Eq. (\ref{eq:JTW_angular}) resulting in the well-known $1/E_e$ behaviour. The latter influences the total decay rate and scalar interactions are strongly constrained by the superallowed $0^+\to0^+$ transitions \cite{Hardy2020}. Tensor interactions are constrained in transitions with a strong Gamow-Teller contribution but which simultaneously have a sufficiently well understood nuclear structure component. The latter translates into low-$A$ transitions within isospin multiplets such as the neutron, nuclear mirror decays and $^6$He \cite{Falkowski2020}.

Recoil spectra then probe $b_F$ through the addition of a term proportional to $I_{010}^C$,
\begin{equation}
    \left.\frac{d\Gamma}{dE_f}\right|_{b_F\neq 0} = \frac{d\Gamma}{dE_f} + b_Fm_eI_{010}^C
\end{equation}
From the results of Fig. \ref{fig:I_shapes} and App. \ref{app:integrals} we may see that there is no particular region in the spectrum that is supremely sensitive to $b_F$. The Coulomb interaction, however, enhances the signal somewhat thanks to the logarithmic divergence in $I_{11-1}$ a $\beta\to 0$.

\subsection{CKM unitarity}

Nuclear beta decay probes the up-down quark mixing matrix element, $V_{ud}$, through the lifetime. If the leading matrix element is sufficiently well-known, it can be extracted from the following master formula
\begin{equation}
    |V_{ud}|^2\mathcal{F}t\left[1+\frac{f_A}{f_V}\rho^2\right] = \frac{K}{G_F^2(1+\Delta_R^V)(M_F^0)^2}
    \label{eq:Vud_master_eq}
\end{equation}
where $K = 8120.274(4)$, $f_A/f_V \approx 1$ is the ratio of axial vector and vector phase-phase integrals, $\Delta_R^V$ is the nucleon-level inner radiative correction, $M_F^0$ is the Fermi matrix element in the isospin limit and $\mathcal{F}t$ is the corrected $ft$ value \cite{Severijns2023}.

For transitions where $M_F^0$ is known sufficiently well, $V_{ud}$ may be extracted given that $\rho$ is known. For mirror decays (including the neutron), one has $M_F^0=1$ and $\rho$ may be experimentally extracted from an angular correlation determination, $X \in \{a_{\beta\nu},A_\beta,B_\nu,\ldots\}$. The sensitivity of different isotopes has been studied before \cite{Hayen2024, Vanlangendonck2022} and is usually expressed through the `enhancement' factor, $\mathcal{E}$, for which $\delta X/X = \mathcal{E}\delta\rho/\rho$. The well-known case of $^{19}$Ne, for example, provides on paper over an order of magnitude increase in precision on $\delta\rho/\rho$ from a measurement of either $a, A$. Whether or not this is a useful metric depends on what kind of uncertainties ultimately dominate the result and how it is determined.

From a recoil spectrum, $a_{\beta\nu}$ may be directly determined as it has an $\mathcal{O}(1)$ influence on the shape of the spectrum. At tree level, one has
\begin{equation}
    a_{\beta\nu}^0 = \frac{1-\rho^2/3}{1+\rho^2}
\end{equation}
so that from a recoil spectrum measurement alone one can extract $|\rho|$. As the recoil spectrum shape contains only a piece that is directly proportional to $a_{\beta\nu}$, however, the relevant sensitivity to $\rho$ is rather $a\times \delta a/a = \delta a$ directly. This significantly shifts the relative sensitivities of different isotopes as a consequence. For $^{19}$Ne, for example, the large sensitivity to $\rho$ originates because of a near cancellation as $a_{\beta\nu}^{19} \approx -0.04$.

Finally, sensitivity to $V_{ud}$ is inversely proportional to $|\rho|$ as is clear from Eq. (\ref{eq:Vud_master_eq}). An compromise then consists in a moderate sensitivity to $\rho$ from $a_{\beta\nu}$ for a small $|\rho|$. This situation changes, however, when one performs a more general fit including Fierz interference terms, as will be discussed in Sec. \ref{sec:statistical_sensitivity}.

\subsection{Sterile neutrinos}
So far we have treated the emitted neutrino as massless with current upper limits at the eV scale \cite{Aker2022}. There is significant scientific interest for a so-called sterile neutrino over a broad mass range, which is a Standard Model singlet under its gauge groups and interacts only with the active family through mass mixing. It is therefore interesting to see what its effects would be on the recoil spectrum. As we will see, these will be relatively minor unless additional variables are measured. A similar conclusion holds for additional exotic particle emission, as will be detailed in a future work.

We may easily generalize the previous results for a finite neutrino mass by performing the following replacements
\begin{equation}
    E_0 \longrightarrow E_0^{m_\nu} = \frac{m_i^2+m_e^2-(m_f+m_\nu)^2}{2m_i}
\end{equation}
in the endpoint energy and decay rate, while the kinematic bounds become
\begin{equation}
    E_e^{\mp,m_\nu}(E_f) = \frac{\Delta_f(s_l+m_e^2-m_\nu^2)\mp p_f\sqrt{\lambda(s_l, m_e^2,m_\nu^2)}}{2s_l}
\end{equation}
with $\Delta_f = m_i-E_f$, $s_l = (p_i-p_f)^2=\Delta_f^2-p_f^2$ and $\lambda$ the K\"all\'en function,
\begin{equation}
    \lambda(s_l, m_e^2, m_\nu^2) = [s_l-(m_e^2+m_\nu^2)][s_l-(m_e^2-m_\nu^2)]
\end{equation}

The recoil spectrum expression of Eq. (\ref{eq:recoil_integral}) remains unchanged as long as the bounds are properly adjusted. A sterile state couples through some admixture, $|U_{e4}|$, where the subscript denotes the coupling between the active electron neutrino and the sterile 4th mass state. Taking the active neutrinos as massless (though it is easily generalized in the same way), the decay rate becomes
\begin{equation}
    \frac{d\Gamma_{tot}}{dE_f} = (1-|U_{e4}|^2)\frac{d\Gamma}{dE_f}+|U_{e4}|^2\frac{d\Gamma_{m_\nu}}{dE_f}
\end{equation}

For $m_\nu \ll m_e$, changes to the spectrum are quadratic only. This can be seen directly from the behaviour of the K\"all\'en function. Near the endpoint, the recoil spectrum obtains a linear sensitivity on $m_\nu$, since
\begin{equation}
    s_{l,min} = (m_e+m_\nu)^2 = m_e^2+2m_em_\nu+m_\nu^2,
\end{equation}
but the change in recoil kinetic energy is strongly reduced because of the nuclear mass, 
\begin{align}
    \delta T_R &\approx \frac{m_em_\nu}{M} \nonumber\\
    &\approx \frac{0.55}{A} \text{eV}\left(\frac{m_\nu}{1~\text{keV}}\right)
\end{align}
The available phase space between the spectral endpoints with a massless and massive neutrino can be approximated as
\begin{equation}
    F_R \simeq \frac{p_{0}(W_0^2+ap_0^2)}{3f_0(W_0)}\left(\frac{m_\nu}{m_e}\right)^3
\end{equation}
with $W_0=E_0/m_e$, $p_0 = \sqrt{W_0^2-1}$ and $f_0$ the dimensionless phase space integral. For $Q\ll m_e$, this can be approximated as
\begin{equation}
    F_R \simeq \frac{35}{16}\left(\frac{m_\nu}{Q}\right)^3
\end{equation}
whereas for $Q\gg m_e$
\begin{equation}
    F_R \simeq 10(1+a_{\beta\nu})\frac{m_\nu^3}{m_eQ^2}
\end{equation}
Relative to the $\beta$ spectrum fraction of $F_E \simeq 10(m_\nu/Q)^3$ then, the recoil spectrum has some benefit for $Q\gg m_e$. The cost is obvious, however, as the energy difference between the massless and massive emission drops with $Q^2$. In general,
\begin{equation}
    \frac{\delta T_R}{T_R^{max}} \simeq \frac{2m_em_\nu}{Q(Q+2m_e)} = \left\{\begin{array}{lc}
        m_\nu/Q, & Q \ll m_e \\
        2m_em_\nu/Q^2, & Q \gg m_e 
    \end{array}\right.
\end{equation}
The limiting factor is the obtainable energy resolution and additional broadening features in the experiment. Performing a kink-type analysis as is done in $\beta$ spectroscopy in principle has a similar sensitivity in the recoil case, be it for a much larger mass range due to the compression with a factor $M^{-1}$.

Substantially more information can be obtained with the detection of the outgoing $\beta$ particle, where now a full Dalitz-type analysis can be performed. Decays with a massive neutrino emission populate regions of phase space forbidden in the three-body massless case. Should additionally the relative direction of the recoil and $\beta$ momenta be known, one may construct a missing mass variable,
\begin{align}
    m^2_{miss} &= (p_i-p_f-p_e)^2 \nonumber \\
    &= (m_i-E_f-E_e)^2-|\bm{p}_e+\bm{p}_f|^2
\end{align}
which is $\delta(m_{miss}^2- m_\nu^2)$ for a three body decay, and a broad continuum for a 4-body decay, $m_{miss}^2=(p_\nu+p_X)^2$. With its first demonstration of a Dalitz distribution measurement of the neutron \cite{Gonzalez2026}, Nab lends itself well to additional searches such as this. 

Recently, an interesting idea was proposed by looking at the `$\beta$ startpoint' \cite{Kodroff2026}, i.e. at the low end of the sum of the $\beta$ particle energy and recoiling nucleus. The lowest visible energy emitted occurs at critical point \textit{B} in Fig. \ref{fig:Dalitz_pedagogical}, which as we mentioned depends quadratically on the neutrino mass,
\begin{equation}
    \left.T_e+T_R\right|^{min} = \frac{Q^2-m_\nu^2}{2(M_f+Q)}
    \label{eq:T_e_T_R_min}
\end{equation}
Even though the sensitivity is reduced by a factor $1/M_f$, the region below this energy is free from Standard Model signals so that it may be used to probe sterile neutrinos. In the original work \cite{Kodroff2026} one considered $^{32}$P as the ideal candidate. This was taken as a compromise due to the unexplained low energy excess found in transition edge sensors and related sensors, but does not apply, for example, to superconducting tunnel junction detectors. Therefore, the sensitivity at low sterile neutrino masses is enhanced for low $M_f$, whereas the window extends up to the $Q$ value. Natural candidates, therefore, are the neutron and $^6$He where the lowest $m_\nu$ probed depends on the achieved detector resolution.

The natural quantity of interest is the fraction of decays with total energy below the Standard Model floor, obtained by setting $m_\nu = 0$ in Eq. (\ref{eq:T_e_T_R_min}), compared to the expected background fraction. This can be calculated directly via
\begin{align}
    \eta_4^<(m_4;Q,M_f,\sigma_E)
&=\frac{|U_{e4}|^2}{\Gamma_0}
\int dT_e\,d\!\cos\theta\,
\frac{d^2\Gamma_4}{dT_e\,d\!\cos\theta} \nonumber \\
&\times \Phi\!\left(
\frac{E_{\rm floor}^{\rm SM}-E_{\rm vis}(T_e,\theta;m_4)}
{\sigma_E}
\right)
\end{align}
where $\Phi$ is the cumulative Gaussian response.

\begin{figure}[ht]
    \centering
    \includegraphics[width=\linewidth]{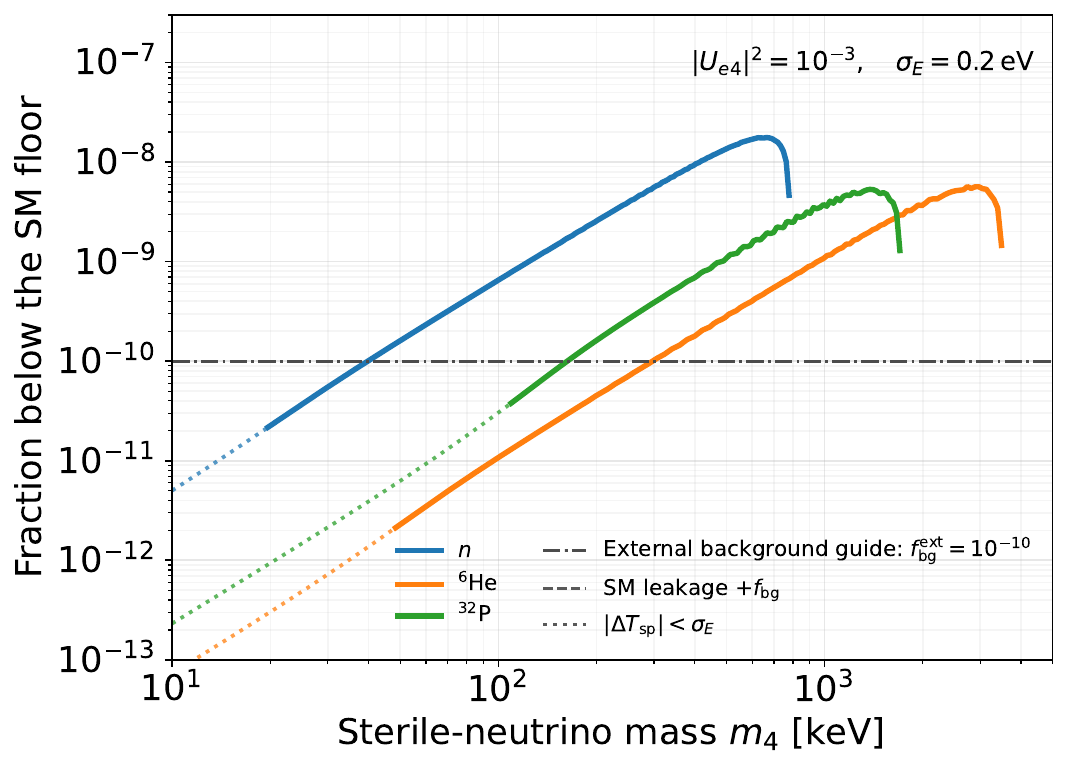}
    \caption{Fraction below the SM floor, $\eta^<_4$, for different nuclear decays as a function of the sterile neutrino mass, $m_4$. The dotted continuation of the lines correspond to when the startpoint shift becomes equal to the detector resolution. The horizontal line corresponds to a background fraction that is imposed by the experimental situation.}
    \label{fig:sterile_below_SM}
\end{figure}

Figure \ref{fig:sterile_below_SM} shows the fraction $\eta_4^<$ as a function of sterile neutrino mass for different nuclear decays. The maximum fraction for the three different decays is of order $|U_{e4}|^2 \times 10^{-5}$. For large $m_4$, the fraction decreases due to the vanishing phase space, while for low masses for overlap with the SM floor becomes increasingly large. The sensitivity window is therefore in the hundreds of keV to a few MeV in $m_4$. Nevertheless, the potential for a background-free measurement sets it apart from, e.g., searches in electron capture \cite{Friedrich2021}.

\begin{figure}[ht]
    \centering
    \includegraphics[width=\linewidth]{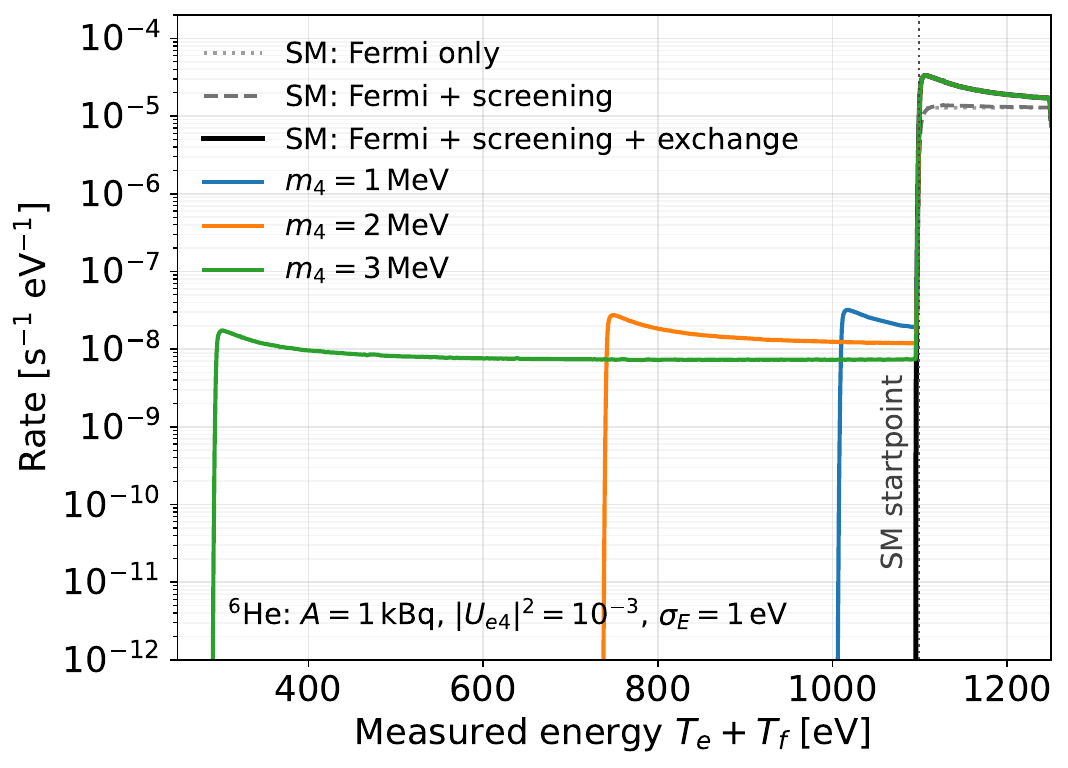}
    \caption{Total energy spectral prediction of the $^6$He $\beta^-$ decay for different sterile neutrino masses. We explicitly show the influence of the Fermi function and exchange corrections, which enhance the low energy rate. }
    \label{fig:6He_startpoint_corrections}
\end{figure}

The fraction $\eta^<_4$ is strongly influenced by the behaviour of the $\beta$ spectrum at low energies. This was not explicitly taken into account in the original paper \cite{Kodroff2026}. It is for this reason that $\beta^-$ decays are vastly more sensitive than $\beta^+$ decays, as the Fermi function strongly influences the spectral density. Similarly, enhancements due to exchange corrections, for example, can significantly increase the proportion. We therefore show a more elaborate version in Fig. \ref{fig:6He_startpoint_corrections}.

\section{Statistical sensitivity}
\label{sec:statistical_sensitivity}
Direct detection of the recoil spectrum has a clear advantage in, for example, direct detection of the $\beta$-$\nu$ angular correlation through a single observable. Because of its large mass relative to the leptons, the spectrum shape is strongly dependent on $a_{\beta\nu}$, and its influence is present throughout the entire range. In order to estimate the statistical sensitivity we construct the Fisher information matrix for a `typical' scenario in which one fits for $a_{\beta\nu}, b_F$ separately or simultaneously. From this we may easily extract the sensitivity to $V_{ud}$ and/or $\epsilon_T$ for a selection of isotopes. We extend the Fisher analysis by adding nuisance parameters such as detector calibration and resolution uncertainties, and unwanted backgrounds to demonstrate when sensitivity is lost. Finally, to optimize the total sensitivity, we consider $\beta$-asymmetry measurements in the same isotopes to see which can provide orthogonal information.

In all of these we construct the Fisher information matrix assuming poisson statistics in each of the bins. For bin $k$, the predicted mean number of counts in that bin is written as $\mu_k(\bm{\theta})$ where $\bm{\theta} = (a, b, \bm{\eta})$ is the set of parameters of interest $\psi \equiv(a, b)$ and nuisance parameters $\bm{\eta}$. The Fisher information matrix is then
\begin{equation}
    \mathcal{I}_{ij} = \sum_{k}\frac{1}{\mu_k}\frac{\partial \mu_k}{\partial \theta_i}\frac{\partial\mu_k}{\partial \theta_j}
    \label{eq:Fisher_I_poisson}
\end{equation}
For our final results we will additionally report the statistical sensitivity in the more meaningful impact on $V_{ud}$ and $\epsilon_T$. This can be directly obtained from the Jacobian
\begin{equation}
    \mathcal{J}^{(a,b)} = \left.\frac{\partial(a,b)}{\partial(V_{ud},\epsilon_T)}\right|_{\text{SM}} = \left(\begin{array}{cc}
    \frac{8K^\prime}{3(1+\rho^2)^2V_{ud}^2}  &  0\\
    0     & \kappa^T(\rho)
    \end{array}\right)
    \label{eq:Jacobian_ab_Vudeps}
\end{equation}
where $K^\prime = K/[G_F^2(M_F^0)^2(1+\Delta_R^V)\mathcal{F}t]$, we have assumed $f_A/f_V = 1$ and defining
\begin{equation}
    \kappa^T(\rho) = \pm2\gamma\frac{\rho^2}{1+\rho^2}\chi^T
\end{equation}
with $\chi^T$ the proportionality constant to $\epsilon_T$ depending on the chosen convention. Using Ref. \cite{Falkowski2020}, one has $\chi^T = 4$. The off-diagonal elements are zero in the Standard Model limit as $\epsilon_T|_{SM}=0$. Introducing additional nuisance parameters simply renders
\begin{equation}
    \mathcal{J}^{(\psi,\eta)} = \left(\begin{array}{cc}
    \mathcal{J}^{\psi}     &  0\\
    0     & \bm{1}_\eta
    \end{array}\right)
\end{equation}
and we obtain the final Fisher information matrix as
\begin{equation}
    \mathcal{I}^{\phi} = \mathcal{J}^T\mathcal{I}^\theta\mathcal{J} = \left(\begin{array}{cc}
        \mathcal{I}_{\phi\phi} &  \mathcal{I}_{\phi\eta}\\
        \mathcal{I}_{\eta\phi} & \mathcal{I}_{\eta\eta}
    \end{array}\right)
\end{equation}
where $\phi = (V_{ud},\epsilon_T,\bm{\eta})$. Marginalizing over nuisance parameters uses the Schur complement resulting in
\begin{equation}
    \mathcal{I}_{\phi|\eta} = \mathcal{I}_{\phi\phi}-\mathcal{I}_{\phi\eta}\mathcal{I}_{\eta\eta}^{-1}\mathcal{I}_{\eta\phi}.
\end{equation}
The per-event information is defined as
\begin{equation}
    \mathcal{I}^{(1)} \equiv \frac{\mathcal{I}}{N}
\end{equation}
From the Cramer-Rao bound we obtain
\begin{equation}
    \text{Cov}(\psi) \simeq (\mathcal{I}_{\psi|\eta})^{-1}
\end{equation}
We note that if all uncertainties are statistical, $\mathcal{I}^{(1)}$ is independent of $N$ and we recover the usual 

\begin{equation}
    \sigma_i = \frac{1}{\sqrt{\mathcal{I}_{i|\text{all}}}} \longrightarrow \frac{1}{\sqrt{N}\sqrt{\mathcal{I}_{i|\text{all}}^{(1)}}}
\end{equation}
behaviour, while some systematic uncertainties discussed in Sec. \ref{sec:nuisance parameters} can cause it to decrease more slowly or bottom out.

If there are only two parameters, the total uncertainty depends on the correlation with each other through
\begin{equation}
    \mathcal{I}_{a|b} = \mathcal{I}_{aa}(1-c^2_{ab}) = \frac{1}{\sigma_{a}^2}
\end{equation}
where $c_{ab} = \mathcal{I}_{ab}/\sqrt{\mathcal{I}_{aa}\mathcal{I}_{bb}}$ with $r_{ab} = \text{Cov}(a,b)/(\sigma_a\sigma_b) = -c_{ab}$ the correlation coefficient.

\subsection{Ideal Fisher information}
In the simplest case, an ideal experiments fits only for three parameters: $a, b$ and a total normalization $s$, i.e. $\bm{\theta} = (a, b, s)$. We may write $\mu_k(\bm{\theta})$ as
\begin{equation}
    \mu_k(\bm{\theta}) = s f_k(a,b), \quad f_k=S_k+aA_k+bB_k
\end{equation}
where $S_k, A_k, B_k$ are the evaluated $I_{s,a,b}$ for bin $k$, respectively. We may construct the Fisher information matrix analytically from Eq. (\ref{eq:Fisher_I_poisson}) around a reference point $\bm{\theta}_0 = (s_0,a_0,b_0)$, so that $f_k^0 \equiv S_k+a_0A_k+b_0B_k$. We may immediately perform the Schur complement to marginalize the influence of the normalization to obtain
\begin{equation}
    \mathcal{I}_{(i,j)|s} = \frac{N}{\sum_kf_k^0}\left[\sum_k\frac{i_kj_k}{f_k^0}-\frac{\sum_ki_k\sum_{k}j_k}{\sum_kf_k^0}\right]
\end{equation}
with $N$ the total number of counts in the fit window.

Table \ref{tab:ideal-statistical-sensitivity} shows the results for all mirror isotopes for $A<40$ together with $^6$He and $^{18}$F. We assume one fits the spectrum between $10\%$ and $90\%$ of the maximal kinetic energy. If one of the parameters is kept fixed, the relevant quantity is simply $\mathcal{I}_{ii}$ rather than $\mathcal{I}_{i|j}$. This would be the case for pure transitions such as $^6$He or $^{18}$F where one would use the Standard Model value for $a=-1/3$.

\begin{ruledtabular}
\begin{table}[t]
\centering
\caption{
Ideal statistical sensitivity coefficients,
$C_x \equiv \sqrt{N}\,\sigma_x
=1/\sqrt{\mathcal I_{xx}^{(1)}}$,
with the other physics parameter held fixed.
The dash indicates that $V_{ud}$ cannot be extracted from a pure
Gamow--Teller transition.
}
\label{tab:ideal-statistical-sensitivity}
\begin{tabular}{lccccc}
Isotope
& $C_a$
& $C_b$
& $C_{V_{ud}}$
& $C_{\epsilon_T}$
& $r_{ab}$ \\
\midrule
$n$                  & 2.77                & 1.55 & 0.893 & 0.116  &  0.835   \\
$^{3}\mathrm{H}$     & $1.19\times10^{2}$ & 1.03 & 38.2  & 0.0789 &  0.99991 \\
$^{6}\mathrm{He}$    & 2.25                & 3.61 & --    & 0.226  & $-0.987$ \\
$^{18}\mathrm{F}$    & 3.24                & 1.38 & --    & 0.0861 & $-0.182$ \\
\midrule
$^{11}\mathrm{C}$    & 2.72 & 1.65 & 0.876 & 0.285 & 0.965 \\
$^{13}\mathrm{N}$    & 2.39 & 1.76 & 0.772 & 0.462 & 0.978 \\
$^{15}\mathrm{O}$    & 2.10 & 2.10 & 0.677 & 0.461 & 0.931 \\
$^{17}\mathrm{F}$    & 2.32 & 2.22 & 0.747 & 0.221 & 0.839 \\
$^{19}\mathrm{Ne}$   & 2.26 & 2.59 & 0.727 & 0.225 & 0.786 \\
$^{21}\mathrm{Na}$   & 1.98 & 2.65 & 0.637 & 0.492 & 0.823 \\
$^{23}\mathrm{Mg}$   & 1.80 & 2.91 & 0.580 & 0.774 & 0.916 \\
$^{25}\mathrm{Al}$   & 1.96 & 3.20 & 0.631 & 0.506 & 0.770 \\
$^{27}\mathrm{Si}$   & 1.86 & 3.52 & 0.598 & 0.673 & 0.872 \\
$^{29}\mathrm{P}$    & 1.73 & 3.49 & 0.556 & 0.973 & 0.955 \\
$^{31}\mathrm{S}$    & 1.70 & 3.79 & 0.547 & 1.08  & 0.966 \\
$^{33}\mathrm{Cl}$   & 1.48 & 3.68 & 0.477 & 2.56  & 0.994 \\
$^{35}\mathrm{Ar}$   & 1.43 & 3.88 & 0.461 & 3.29  & 0.995 \\
$^{37}\mathrm{K}$    & 1.72 & 4.36 & 0.554 & 1.09  & 0.962 \\
$^{39}\mathrm{Ca}$   & 1.78 & 4.71 & 0.574 & 0.969 & 0.944 \\
\end{tabular}
\end{table}
\end{ruledtabular}

\begin{figure}[ht]
    \centering
    \includegraphics[width=\linewidth]{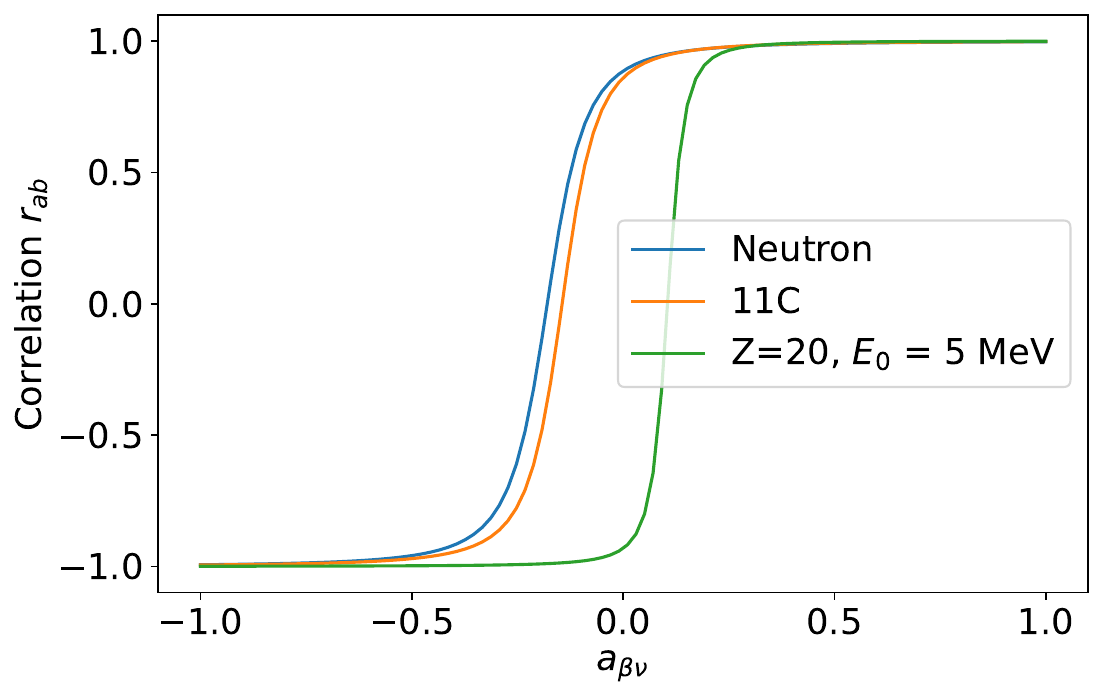}
    \caption{Statistical correlation parameter between $a$ and $b$ as a function of $a$ for different $\beta$ decay transitions, fitted between 10-90\% of the total energy interval. For the neutron, one finds $r_{ab} \approx 0.8$, while for $^{11}$C $r_{ab} \approx 0.965$.}
    \label{fig:correlation_rab}
\end{figure}

One noteworthy thing is the strong correlation between $a$ and $b$ for most isotopes. Taking $^{11}$C as an example, one obtains a correlation parameter of $r_{ab} \approx 0.965$ so that the marginalized uncertainties are a factor $(1-r_{ab}^2)^{-1/2}\sim4$ larger than if one of them is kept constant. For reference, we have plotted the correlation coefficient $r_{ab}$ for the neutron, $^{11}$C and an $A=40, E_0 = 5$ MeV decay as a function of $a$ in Fig. \ref{fig:correlation_rab}.

Except for very specific values of $a$ the correlation quickly grows to $r_{ab} \to \pm 1$. The eigenvectors of the Fisher information align more and more with $a\pm b$ as a result. This will be a central feature of Sec. \ref{sec:orthogonal_observables}, where we investigate for which isotopes one may obtain much more information through a measurement of $A_\beta$ with orthogonal constraints.

\subsection{Nuisance parameters: calibration, resolution, background}
\label{sec:nuisance parameters}

In a real experiment, there will be a number of nuisance parameters due to the energy calibration, backgrounds to be subtracted, and so on. For Fierz measurements in direct $\beta$ spectroscopy it is well-known, for example, that the uncertainty on the calibration must often be at least an order of magnitude smaller than the precision aimed-for on $b_F$. We now investigate whether a similar effect occurs in the recoil spectrum. We may use the Fisher machinery of the previous section and construct a simplified detector model,
\begin{equation}
    \lambda(x) = N_S \int dE s(E;a,b)R(x|E, \bm{\eta}_R) + N_B\mathcal{B}(x,\bm{\eta}_B)
\end{equation}
where $\mathcal{B}$ is a linear background, and $R$ is a simplified detector response function
\begin{equation}
    R(x|E,\bm{\eta}_R) = \frac{1}{\sqrt{2\pi\sigma(E,\eta_q)}}\exp\left(-\frac{(x-\bar{x}(E,\bm{\eta}_c))^2}{2\sigma^2(E,\eta_{q})}\right)
\end{equation}
where the calibration is quadratic, i.e.
\begin{equation}
    \bar{x} = \beta_0 + \beta_1u+\beta_2u^2
    \label{eq:calibration_u}
\end{equation}
with $u = (E-E^\star)/E_\text{scale}$. $E^\star$ can be chosen halfway in the analysis window and $E_\text{scale}$ about half the analysis window width, such that $|u| \lesssim 1$. A rescaled calibration like this - rather than taking $\bar{x} = \sum_i c_iE^i$ - significantly reduces correlations between parameters.

All of these parameters are typically obtained through a separate data taking campaign and therefore contain prior information. As these are usually statistically independent datasets, the likelihoods multiply and the Fisher matrices simply add. Given a certain covariance matrix, $V_{\eta\eta}$ from, e.g., a previous calibration campaign, one may construct the full Fisher information matrix by
\begin{align}
    \mathcal{I} &= N_S \mathcal{I}^{(1)}+V^{-1}_{\eta\eta} \\
    &= \left(\begin{array}{cc}
        \mathcal{I}_{\psi\psi} & \mathcal{I}_{\psi\eta} \\
        \mathcal{I}_{\eta\psi} & \mathcal{I}_{\eta\eta} + V_{\eta\eta}^{-1}
    \end{array}\right)
\end{align}
This result offers a clean interpretation. Strong prior constraints on nuisance parameters are such that $V^{-1}_{\eta\eta} \gg \mathcal{I}_{\eta\eta}=N_S\mathcal{I}_{\eta\eta}^{(1)}$. This is particularly important for nuisance parameters mimic the $a$ or $b$ signal. We will therefore proceed in a two-step fashion: first, we will look at the Fisher correlation matrix $\mathcal{I}_{\eta\eta}^{(1)}$ as it shows which nuisance parameters should be constrained most. This, however, neglects correlations. Second, we will then look at a realistic exposure of $N_S = 10^8$ and perform the proper marginalization of the full Fisher matrix for some level of precision and correlation.

Let us look at the simpler one-parameter nuisance case. For a nuisance parameter $j$ let
\begin{equation}
    H_j = N_S \mathcal{I}^{(1)}_{jj}, \quad p_j = \frac{1}{\sigma_{j,\text{prior}}^2}
\end{equation}
for $p_j$ some prior knowledge. The approximate information loss is then
\begin{equation}
    L_{\psi j} = \left(c^\text{shape}_{\psi j}\right)^2\frac{H_j}{H_j+p_j}, \quad c_{\psi j}^\text{shape} = \frac{\mathcal{I}_{\psi j}^{(1)}}{\sqrt{\mathcal{I}_{\psi\psi}^{(1)}\mathcal{I}_{jj}^{(1)}}}
\end{equation}
where the information reduces as $\sqrt{1-L_{\psi j}}$. If we want the information loss to be below some threshold $\delta$, that then requires
\begin{equation}
    \sigma_{j,\text{prior}} < \frac{1}{\sqrt{H_j(c_{\psi j}^2/\delta-1)}}.
\end{equation}

\begin{figure}
    \centering
    \includegraphics[width=\linewidth]{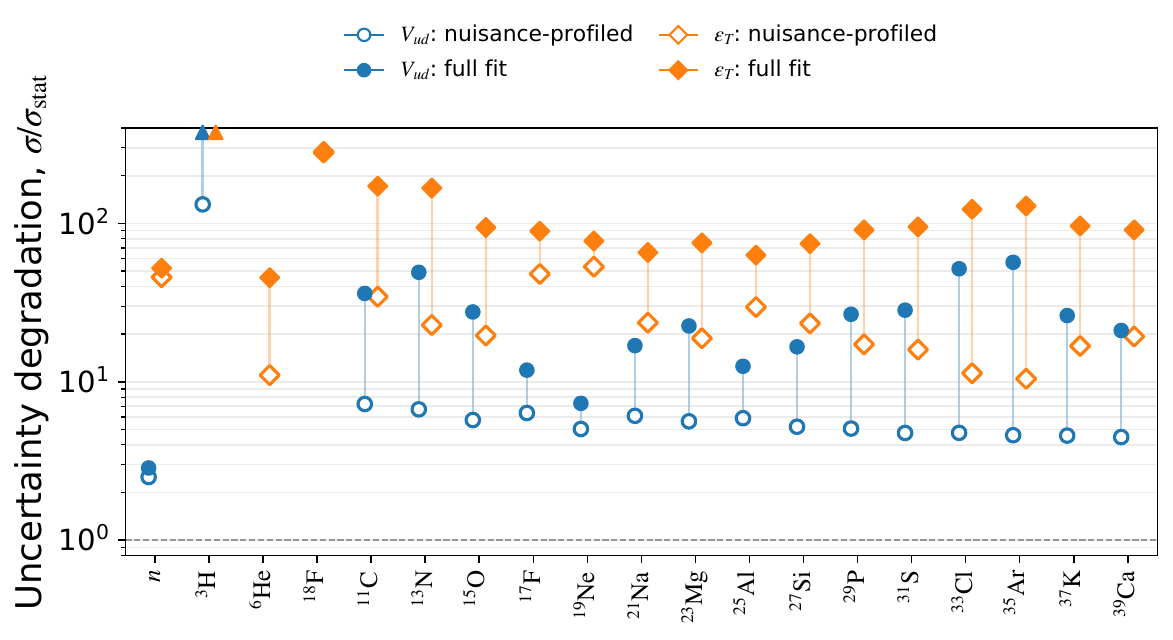}
    \caption{Uncertainty degradation for different isotopes in $\epsilon_T$ and $V_{ud}$ due to profiling over the nuisance parameters due to detector response and background, versus a full fit extraction of $(V_{ud}, \epsilon_T)$ together.}
    \label{fig:uncertainty_degradation}
\end{figure}

In order to introduce an explicit example we perform a $z$-scoring of all parameters. Given representative values for the precision by which parameters are known, this results in a map of how each parameter affects an extraction of $a$ and $b$. We may compare that to the uncertainty one obtains in the purely statistical case of the previous section and compare different isotopes' sensitivity. This is shown in Fig. \ref{fig:uncertainty_degradation}.

\begin{figure}
    \centering
    \includegraphics[width=\linewidth]{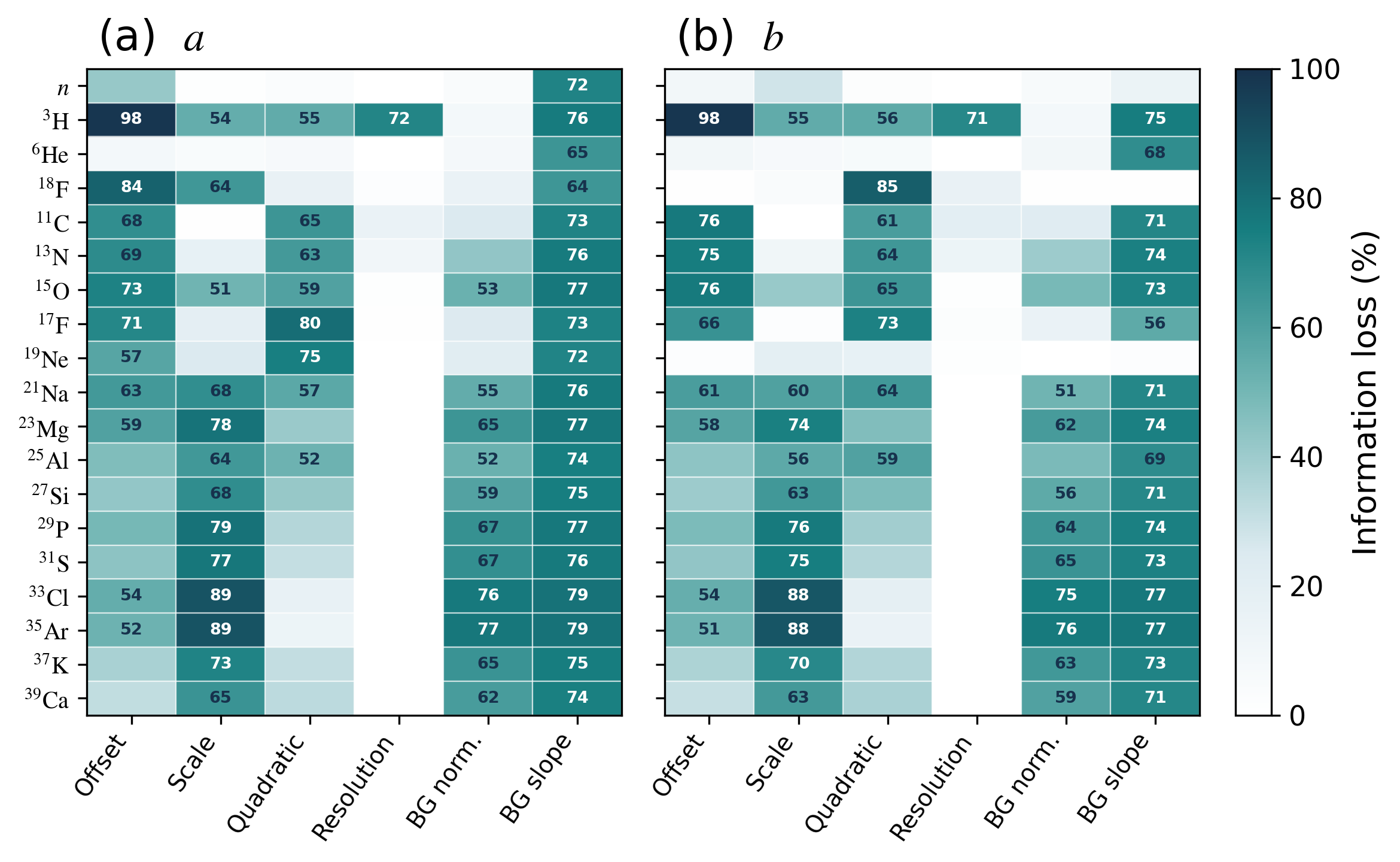}
    \caption{Single-nuisance information loss expressed as a percentage for either $a$ or $b$ for different physics cases.}
    \label{fig:single_nuisance_information_loss}
\end{figure}

Figure \ref{fig:single_nuisance_information_loss} shows the effect of individual nuisance parameters on the information loss of either $a$ or $b$. Here, the fit is performed only in the region of interest, defined as the 10-90\% interval. Constraints on fit parameters due to external measurements are added appropriately as additional information through $V^{-1}_{\eta\eta}$. In doing so, we do have to make a choice on the correlation between different parameters. Energy calibration parameters, for example, will often be correlated depending on the functional form. It was partly for this reason we have chosen Eq. (\ref{eq:calibration_u}), where for the exercise we consider the parameters to be completely decorrelated. This can obviously be easily modified for an actual experimental analysis. 

For all cases, the information loss is strong from the background slope. In most scenarios, this can relatively easily be studied with sufficiently high precision. More interesting is the influence of the calibration. Higher-mass mirror decays have a higher endpoint energy, so that for the first half of the decays the quadratic component of the calibration is dominant. As the endpoint grows, the uncertainty in the linear calibration terms couples more strongly with the physics signal that it needs to be constrained to much higher precision.

\subsection{Orthogonal observables}
\label{sec:orthogonal_observables}

Finally, we consider the added value of an $A_\beta$ measurement in the same isotope. This is motivated by the findings at the beginning of this section where the recoil spectrum from most mirror decays results in a highly correlated extraction of $a$ and $b$. While this is not intrinsically bad, it does mean that external constraints are necessary for competitive extractions of either one.

The leading order $\beta$-asymmetry can be written as
\begin{equation}
    A_\beta = \frac{\pm \rho^2/(J+1)+2\sqrt{\frac{J}{J+1}}\rho}{1+\rho^2}
\end{equation}
where the sign denotes $\beta^\mp$ decay. In the presence of a non-zero Fierz term, we may construct the per-bin asymmetry as
\begin{equation}
    \tilde{A}_j = \frac{A_0}{1+b_F\langle\frac{m_e}{E_e}\rangle}_j
\end{equation}
where $\langle m_e/E_e\rangle_j$ is the average in bin $j$.

If the experiment does not have calorimetric capabilities, one simply measures an average $\tilde{A}_\beta$. If, on the other hand, some energy sensitivity is obtained one may attempt to separately extract $\rho$ and $b_F$. In order to look at the additional value of a $\beta$-asymmetry, we may construct the Jacobian in the same fashion as \ref{eq:Jacobian_ab_Vudeps} and put everything in the same basis. Then, as they are independent measurements the Fisher information matrices simply add. More specifically, we have
\begin{equation}
    F_A^{(V_{ud}, \epsilon_T)} = \sum_i G_i^T C_{A,i}^{-1}G_i
\end{equation}
where 
\begin{equation}
    G_j = \left(dA/dV_{ud} \quad -A_0\kappa^T\langle m_e/E_e\rangle_j\right)
\end{equation}
for a covariance matrix $C_A$. We may write down the variance of the bin-by-bin asymmetry, $R_j$, as
\begin{equation}
    \textrm{Var}(R_j) = \frac{1-R_j^2}{N_j}, \quad R_j = \alpha_j \frac{A_0}{1+b_Fm_e/E_j}
\end{equation}
with $\alpha_j$ the combination of polarization, detection solid angle, etc. Using this notation, the Fisher matrix may we written as
\begin{equation}
    F^A_{\alpha\beta} = \sum_j \frac{N_j}{1-R^2_j}\frac{\partial R_j}{\partial\theta_\alpha}\frac{\partial R_j}{\partial\theta_\beta}.
\end{equation}
From this result, one can see that the Fisher information scales \textit{quadratically} with the polarization (contained in $\alpha_j$) for small-to-medium $R_j$. Any uncertainty on the polarization enters as a component $1/\sigma_P^2$ in the Fisher matrix as usual.

Assuming we may perform a similar $\beta$ decay experiment and fit the resulting asymmetry over the 10-90\% interval of the $\beta$ spectrum, we look at the information gain in an idealized case for each isotope. The results are listed in Table \ref{tab:mirror-bigA-complementarity}.

\begin{ruledtabular}
\begin{table}[tb]
\caption{Joint recoil--$A_\beta(E_e)$ benchmark for
$N_{\rm recoil}=N_A=N_0=10^8$ accepted events. The coefficients are
$C(N_0)=\sqrt{N_0}\,\sigma$; fixed external detector constraints make them
benchmark-normalized rather than exactly exposure independent. The signed
$r_{Ab}^{(A)}$ is the covariance correlation of the energy-resolved
asymmetry-only fit in $(A,b)$. The tensor gain is
$G_T=\sigma_{\epsilon_T}^{\rm recoil}/\sigma_{\epsilon_T}^{\rm joint}$.
Pure Gamow--Teller decays do not determine $V_{ud}$; $^6$He has $J_i=0$
and no beta asymmetry.}
\centering
\small
\setlength{\tabcolsep}{7pt}
\begin{tabular}{lrrrr}
Isotope & $C_{V_{ud}}^{a+A}$ & $C_{\epsilon_T}^{a+A}$ & $r_{Ab}^{(A)}$ & $G_T$ \\
\midrule
$n$                    & 1.01                    & 3.89  & $-0.989$ & 1.56                    \\
$^{3}\mathrm{H}$       & $2.61\times10^{3}$      & 131   & $-1.00$  & $2.62\times10^{3}$      \\
$^{6}\mathrm{He}$      & --                      & 2.49  & --       & 1.00                    \\
$^{18}\mathrm{F}$      & --                      & 1.87  & $+0.992$ & 12.9                    \\
$^{11}\mathrm{C}$      & 6.18                    & 1.24  & $-0.987$ & 39.6                    \\
$^{13}\mathrm{N}$      & 5.08                    & 3.50  & $-0.983$ & 22.1                    \\
$^{15}\mathrm{O}$      & 1.77                    & 3.88  & $+0.977$ & 11.2                    \\
$^{17}\mathrm{F}$      & 4.82                    & 0.664 & $+0.977$ & 29.8                    \\
$^{19}\mathrm{Ne}$     & 1.07                    & 13.3  & $-0.972$ & 1.31                    \\
$^{21}\mathrm{Na}$     & 2.15                    & 3.23  & $+0.970$ & 9.98                    \\
$^{23}\mathrm{Mg}$     & 2.12                    & 5.09  & $-0.966$ & 11.4                    \\
$^{25}\mathrm{Al}$     & 2.57                    & 2.58  & $+0.965$ & 12.4                    \\
$^{27}\mathrm{Si}$     & 2.58                    & 2.95  & $-0.962$ & 17.0                    \\
$^{29}\mathrm{P}$      & 1.19                    & 7.50  & $+0.961$ & 11.8                    \\
$^{31}\mathrm{S}$      & 2.46                    & 7.40  & $-0.959$ & 13.9                    \\
$^{33}\mathrm{Cl}$     & 0.661                   & 20.5  & $-0.958$ & 15.3                    \\
$^{35}\mathrm{Ar}$     & 0.438                   & 26.2  & $+0.957$ & 16.2                    \\
$^{37}\mathrm{K}$      & 1.84                    & 5.71  & $-0.956$ & 18.4                    \\
$^{39}\mathrm{Ca}$     & 1.47                    & 5.18  & $+0.955$ & 17.0                   
\end{tabular}
\label{tab:mirror-bigA-complementarity}
\end{table}
\end{ruledtabular}

Isotopes for which the correlation $r_{Ab}$ is similar to $r_{ab}$ offer little additional information as they probe the same eigenmode of the Fisher matrix. This is the case, for example, for $^{19}$Ne and the resultant information gain is marginal. A case like $^{11}$C, on the other hand, contains quasi-orthogonal constraints from $a$ and $A$ and the resultant uncertainty is decreased by a factor of 40.

\begin{figure}[ht]
    \centering
    \includegraphics[width=\linewidth]{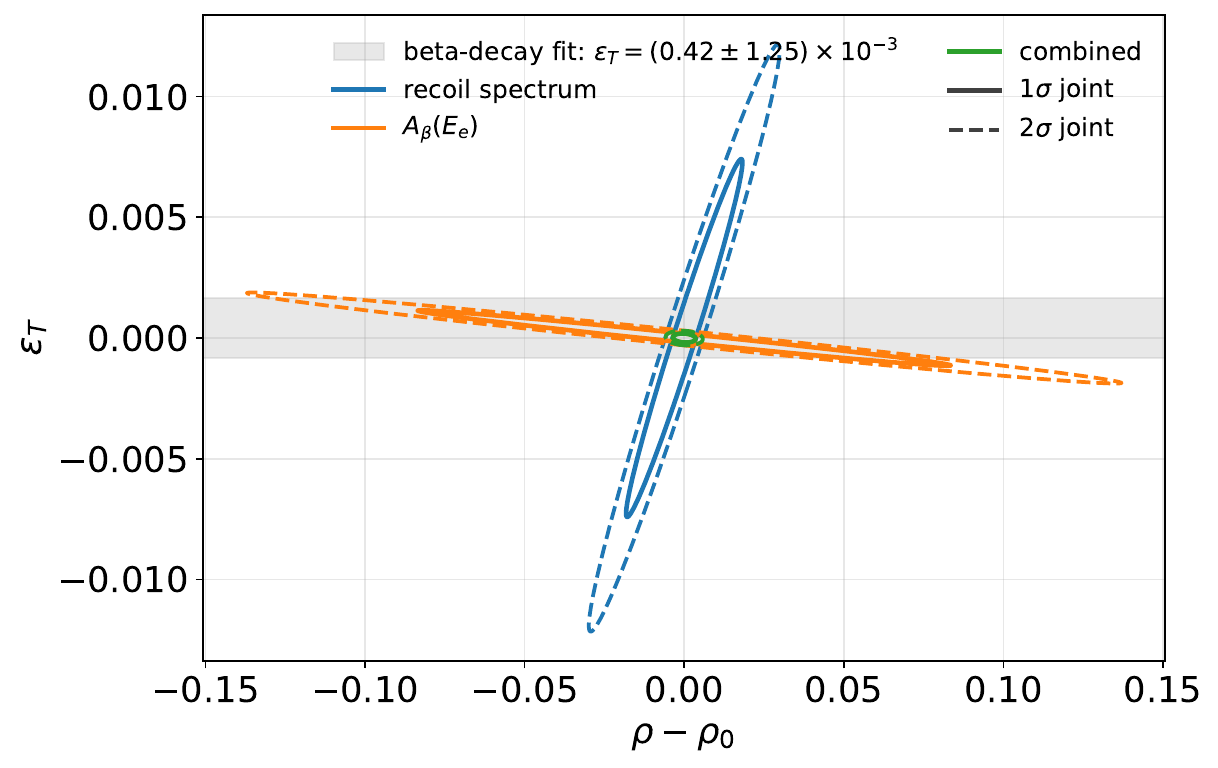}
    \caption{Constraints on $\rho$ and $\epsilon_T$ from a simultaneous determination of $a$ and $A_\beta$ for $^{11}$C in an idealized scenario for $N=10^8$ counts. Additionally, current constraints from global $\beta$ decay fits are shown in the grey band from Ref. \cite{Falkowski2020}.}
    \label{fig:11C_complementarity}
\end{figure}

In order for this to be a viable strategy, one must be able to polarize the initial nucleus. There is not an experimentally tested strategy for for production of polarized nuclei for all of the isotopes listed. Polarized $^{11}$C, for example, has not been demonstrated as of yet. For $^{17}$F, on the other hand, which presents a gain factor of 30 already initial measurements in the 1970s have demonstrated polarization following the $^{16}$O$(d,n)^{17}$F reaction. Preliminary measurements of the $\beta$-asymmetry were performed using low temperature nuclear orientation but no further analysis was developed \cite{Severijns1989}.

\section{Conclusion}

New techniques mean that the measurement of the kinetic energy of the recoiling nucleus following $\beta$ decay is becoming directly accessible. Here, we have presented a complete description of various kinds of Standard Model corrections to allow for a clean extraction of Beyond Standard Model constraints. Noteworthy is the demonstration that the theoretical recoil spectrum is an exceptionally well-known observable, as most final state interactions largely average out, and dominant nuclear structure uncertainties vanish at leading order. We have described radiative corrections in some depth, and separate the contribution of both hard and soft bremsstrahlung photons on the spectrum. Even so, several of these corrections provide corrections to the determination of the beta-neutrino angular correlation at the current or anticipated level of precision. Fortuitously, uncertainties on these corrections are small as their descriptions concern using exact kinematics, radiative corrections due to bremsstrahlung emission, and well-studied final state interactions. As in $\beta$ decay spectroscopy, nuclear structure corrections carry the largest relative uncertainty, but their effects are at least two orders of magnitude smaller in the recoil spectrum rendering them completely negligible.

We have described a number of different, exciting Beyond Standard Model extensions, ranging from exotic current searches, CKM unitarity and sterile neutrino searches. For these scenarios, we constructed a rigorous statistical analysis describing the obtained information and the correlations between both physics and nuisance parameters. In an idealized scenario, recoil spectroscopy with only a million decays provide competitive constraints to the global $\beta$ decay data set. We have additionally shown that the introduction of nuisance parameters typical to a set up like this can significantly modify this precision by substantially reducing the Fisher information. We derived analytical constraints experimentalists can use to identify dominant systematic uncertainties and provide scaling functions to obtain a required level of control.

Finally, we demonstrated that a careful selection of isotopes can provide constraints in shared fits that are up to a factor 40 stronger than the individual results by combining recoil and $\beta$-asymmetry measurements with opposite correlations between the physics parameter. This exercise resulted in a set of isotopes other than those usually identified as go-to choices, such as $^{11}$C and $^{17}$F. Correlations between observables for $^{19}$Ne, for example, mean that the combined statistical sensitivity increases only by 30\%. As such, this presents an opportunity for the community to develop cases based on their total statistical information provided and more effectively probe the discovery potential of new physics at low energies.

\begin{acknowledgements}
    This research is supported by the French National Agency for Research, Agence National de la Recherche. I thank Victor Dumenil for the calculation of the induced tensor matrix element of $^{18}$F. 
\end{acknowledgements}

\appendix

\section{Kinematic integrals}
\label{app:integrals}

For ease of use we provide some analytical results for the parametrized shapes, $I_{mnl}$, as indefinite integrals. The contribution to the recoil spectrum for an energy $E_f$ is then obtained via its evaluation at $E_{e, min/max}(E_f)$. All $I_{mnl}$ for arbitrary $n$ can naturally be obtained through linear combinations of $I_{m0l}$. The Coulomb versions to first order in $\alpha Z$ can be obtained through Eq. (\ref{eq:I_Coul_expansion}).

\begin{subequations}
\begin{align}
    I_{110} &= E_e(E_0/2-E_e^2/3)\\
    I_{010} &= E_e(E_0-E_e/2) \\
    I_{002} &= E_e(E_e^2/3-m_e^2) \\
    I_{020} &= (E_0-E_e)^3/3\\
    I_{101} &= p_e^3/3\\
    I_{10-1} &= p_e/2 \\
    I_{20-1} &= \frac{1}{2}\left(p_eE_e+m_e^2\ln\left(\frac{1+\beta}{1-\beta}\right)\right) \\
    I_{30-1} &= \frac{1}{3}p_e(2m_e^2+E_e^2)\\
    I_{10-2} &= \ln(p_e/m_e)\\
    I_{10-3} &= -1/p_e
\end{align}
\end{subequations}

\bibliography{library}

\end{document}